\documentclass{article}

\usepackage{placeins}
\usepackage{arxiv}

\usepackage[utf8]{inputenc} 
\usepackage[T1]{fontenc}    
\usepackage{hyperref}       
\usepackage{url}            
\usepackage{booktabs}       
\usepackage{amsfonts}       
\usepackage{nicefrac}       
\usepackage{microtype}      
\usepackage{graphicx}
\usepackage{soul}
\usepackage[numbers]{natbib}
\usepackage{doi}
\usepackage{amsmath}
\usepackage{xcolor} 

\title{Bayesian Neural Networks versus deep ensembles for uncertainty quantification in machine learning interatomic potentials}

\author{ \href{https://orcid.org/0000-0000-0000-0000}{\includegraphics[scale=0.06]{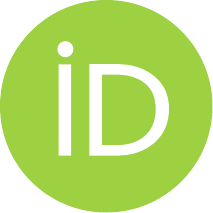}\hspace{1mm}Riccardo Farris}\thanks{Corresponding author: rfarris@ub.edu} \\
Leitat Technological Center\\ Carrer de la Innovació, 2,
08225\\ Terrassa, Spain\\
	Departament de Ci\`{e}ncia de Materials\\
        i Qu\'{\i}mica F\'{\i}sica \& Institut de Qu\'{\i}mica \\
        Te\`{o}rica i Computacional (IQTCUB)\\
        Universitat de Barcelona\\
        Barcelona, Spain\\
        \And
    	\href{https://orcid.org/0000-0000-0000-0000}{\includegraphics[scale=0.06]{orcid.pdf}\hspace{1mm}Emanuele Telari} \\
    	Departament de Ci\`{e}ncia de Materials\\
            i Qu\'{\i}mica F\'{\i}sica \& Institut de Qu\'{\i}mica \\
            Te\`{o}rica i Computacional (IQTCUB)\\
            Universitat de Barcelona\\
            Barcelona, Spain\\
        \And
    	\href{https://orcid.org/0000-0000-0000-0000}{\includegraphics[scale=0.06]{orcid.pdf}\hspace{1mm}Nongnuch Artrith} \\
    	Debye Institute for Nanomaterials Science\\
            Utrecht University\\
             Utrecht, Netherlands\\
        \And
        \href{https://orcid.org/0000-0002-5242-5567}{\includegraphics[scale=0.06]{orcid.pdf}\hspace{1mm}Konstantin M. Neyman} \\
	Departament de Ci\`{e}ncia de Materials\\
        i Qu\'{\i}mica F\'{\i}sica \& Institut de Qu\'{\i}mica \\
        Te\`{o}rica i Computacional (IQTCUB)\\
        Universitat de Barcelona\\
        Barcelona, Spain\\
        ICREA (Institucio Catalana de Recerca i Estudis Avançats)\\ Pg. Lluıs Companys 23\\ 08010 Barcelona, Spain
	\And
	\href{https://orcid.org/0000-0000-0000-0000}{\includegraphics[scale=0.06]{orcid.pdf}\hspace{1mm}Albert Bruix}\thanks{Corresponding author: 
 abruix@ub.edu} \\
	Departament de Ci\`{e}ncia de Materials\\
        i Qu\'{\i}mica F\'{\i}sica \& Institut de Qu\'{\i}mica \\
        Te\`{o}rica i Computacional (IQTCUB)\\
        Universitat de Barcelona\\
        Barcelona, Spain\\
}

\renewcommand{\headeright}{ }
\renewcommand{\undertitle}{ }
\renewcommand{\shorttitle}{}

\hypersetup{
pdftitle={Bayesian Aenet},
pdfsubject={q-bio.NC, q-bio.QM},
pdfauthor={Riccardo Farris},
pdfkeywords={Bayesian Neural Networks, Variational Inference, Deep Ensembles, Uncertainty Quantification, Machine Learning Interatomic Potentials},
}
\date{}
\begin{document}
\maketitle

\begin{abstract}
Neural-network-based machine learning interatomic potentials have emerged as powerful tools for predicting atomic energies and forces, enabling accurate and efficient simulations in atomistic modeling. A key limitation of traditional deep learning approaches, however, is their inability to provide reliable estimates of predictive uncertainty. Such uncertainty quantification is critical for assessing model reliability, especially in materials science, where often the model is applied on out-of-distribution data. Different strategies have been proposed to address this challenge, with deep ensembles (DE) and Bayesian neural networks (BNN) being among the most widely used. In this work, we introduce an implementation of Bayesian neural networks with variational inference in the \texttt{\ae net-PyTorch} framework. To evaluate their applicability to machine learning interatomic potentials, we systematically compare the performance of variational BNNs and deep ensembles on a dataset of 7,815 TiO\textsubscript{2} structures, and further assess the generalisability of our conclusions on the chemically more diverse QM7 dataset of 7,165 small organic molecules. The models are trained on both a full dataset and a smaller subset to assess how variations in data representation influence predictive accuracy and uncertainty estimation. Our findings reveal that, despite its more empirical and less theory-grounded nature and in addition to its lower inference cost and implementation simplicity, Deep Ensembles outperform variational Bayesian Neural Networks on most accuracy and uncertainty-quantification metrics across data regimes, the main exceptions being the calibration error and the energy accuracy on the chemically diverse QM7 dataset, where the BNNs are competitive or better. This analysis provides insights into the strengths and limitations of each approach, offering practical guidance for the development and evaluation of uncertainty-aware machine learning interatomic potentials.

\end{abstract}


\section{Introduction}

In recent years, machine learning interatomic potentials (MLIPs) have emerged as powerful tools to overcome the prohibitive computational cost of \textit{ab initio} methods such as density functional theory (DFT). The most promising MLIPs leverage deep learning architectures, such as feed-forward neural networks~\citep{Behler2007,Artrith2014}, and more recently, graph neural networks~\citep{batatia2024foundationmodelatomisticmaterials}, to learn the mapping between atomic configurations and their corresponding energies and forces from a limited set of reference calculations~\citep{Deringer2019}. These models have proven capable of achieving superior accuracy and transferability, even when trained on relatively limited datasets. This efficiency dramatically expands the accessible length and time scales, enabling the study of larger and more complex systems that would otherwise be infeasible with traditional first-principles approaches~\citep{behler2016perspective, zuo2020performance}.

A critical aspect in the development of such MLIPs is the generation of high-quality datasets for training purposes, due to the high computational cost of first-principles calculations, used as the reference level of theory~\citep{Mueller2020, Artrith2021}. This severely limits the size and diversity of training datasets, posing a major bottleneck in the construction of robust and transferable MLIPs. As a result, MLIP models need to be designed not only for accuracy but also for data efficiency, in order to minimize the need for exhaustive DFT sampling while still maintaining predictive reliability across diverse atomic configurations.
This calls for finding ways to reliably quantify the uncertainty associated with their predictions, preventing misleading results in simulations, and guiding the selection of informative data points for model refinement in active learning scenarios.~\citep{kulichenko2023uncertainty, christiansen2024efficient, kellner2024uncertainty}. 

Standard deep learning architectures are inherently deterministic and do not provide any measure of prediction uncertainty. As a result, the deployment of deep learning-based MLIPs in uncertainty-sensitive applications requires the incorporation of specialized methods for its quantification, such as dropout~\citep{wen2020uncertainty},  conformal prediction~\citep{hu2022robust}, Gaussian  Mixture Model~\citep{zhu2023fast}, and quantile regression~\citep{bilbrey2025uncertainty}. Among the various approaches proposed, deep ensembles (DE), proposed initially in the work from Lakshminarayanan et al.~\citep{Lakshminarayanan2016}, have emerged as one of the most widely adopted techniques~\citep{Scalia2020}, due to their simplicity and straightforward application to more complicated network architectures~\citep{Carrete2023}, and their robust performance~\citep{tan2023single}. In its initial formulation, the ensembling was achieved by training multiple neural networks independently with different initializations on the same dataset, and the variance across their predictions is used as a proxy for model uncertainty. Subsequently, several alternative ensembling strategies have been proposed to compute uncertainty and are nowadays more commonly used, like committee neural networks ~\citep{schran2020committee} and multi-head architectures~\citep{beck2025multi}, which allow to obtain uncertainty estimates, lowering the computational cost of training and forwarding completely different models. Although DE are effective in practice and relatively simple to implement at inference time, they sometimes lead to overconfident predictions in regions of configuration space that are poorly represented in the training data~\citep{Lu2023, Kahle2022}.

In addition to DE, a specialized class of neural networks has emerged to intrinsically offer a measure of their predictions uncertainties by including the Bayesian formalism into the network architecture, the so-called Bayesian Neural Networks (BNNs)~\citep{mackay1992}. BNNs are grounded in the Bayesian probability framework, which treats model parameters, i.e. weights and biases, as random variables with prior distributions~\citep{Jospin2020}. By incorporating Bayes theorem, BNNs place a posterior distribution over the model parameters conditioned on the training data. Predictions are then obtained by sampling from the resulting predictive distribution, which propagates parameter uncertainty and naturally captures uncertainty in the model’s outputs.
Given a dataset $\mathbf{X}, \mathbf{Y}$ where $\mathbf{X}$ is the descriptor matrix and $\mathbf{Y}$ are the reference targets and  the model parameters $\mathbf{w}$, the posterior distribution is computed using Bayes’ theorem:

\begin{equation}
    p(\mathbf{w} \mid \mathbf{X}, \mathbf{Y}) = \frac{p(\mathbf{Y} \mid \mathbf{X},  \mathbf{w}) \, p(\mathbf{w})}{p(\mathbf{Y} \mid \mathbf{X})},
\end{equation}

where $p(\mathbf{w})$ is the prior distribution over the parameters in the parameter space $\Omega$, $p(\mathbf{Y} \mid \mathbf{X},  \mathbf{w})$ is the likelihood of the data given the parameters $\mathbf{w}$, and $p(\mathbf{Y} \mid \mathbf{X})$ is the marginal likelihood or evidence.

Based on this posterior, the predictive distribution for a new input $\mathbf{x}^*$ is given by:

\begin{equation}
     p(\mathbf{y}^* \mid \mathbf{x}^*, \mathbf{X}, \mathbf{Y}) = \int_{\Omega} p(\mathbf{y}^* \mid \mathbf{x}^*, \mathbf{w})\, p(\mathbf{w} \mid  \mathbf{X}, \mathbf{Y})\, d\mathbf{w}
     \label{eq:pred_dist}
\end{equation}

where $\mathbf{y}^* $ is the predicted output, the energy in the case of MLIPs, corresponding to the input features, or descriptor, $\mathbf{x}^*$. The key challenge in this formulation lies in the intractability of the integral in Equation~\ref{eq:pred_dist}, which arises from the high dimensionality and nonlinearity of the parameter space. As a result, approximate inference techniques must be employed to estimate such distribution. A traditional approach to this problem involves sampling from the posterior using Markov Chain Monte Carlo (MCMC) methods, including more efficient variants such as Hamiltonian Monte Carlo (HMC)~\citep{Brooks2011}. However, MCMC-based techniques are computationally expensive and often impractical for high-dimensional deep learning models. As a result, variational inference has become a more widely adopted alternative in Bayesian deep learning~\citep{blundell2015}. Variational inference approximates the true posterior by optimizing over a family of tractable, parameterized distributions, thus enabling scalable and efficient inference at the cost of introducing a variational approximation error~\citep{Jospin2020}.

Nonetheless, inferring the posterior distribution in BNNs remains significantly more complex and computationally demanding than training standard deterministic neural networks. This added complexity constitutes a major practical limitation of BNNs, especially when applied to high-dimensional systems as commonly encountered in materials simulations.

In the present work, we explore the applicability and the performance of different techniques to measure the uncertainty of deep learning-based MLIPs, in particular deep ensembles and Bayesian neural networks trained with variational inference, which we will refer to as variational Bayesian neural networks (VBNN).
Starting from the \texttt{\ae net} library~\citep{Artrith2016, Lpez-Zorrilla2023}, we propose a VBNN implementation, which supports GPU-accelerated training, for MLIPs based on the work of Basora et al.~\citep{Basora2023}, available at~\url{https://github.com/farrisric/bayesaenet}. We employed the Pyro~\citep{bingham2018pyro} probabilistic programming framework, designed for Bayesian inference, along with the TyXe~\citep{Ritter2021} library, which provides tools for converting standard neural networks into Bayesian neural networks. We then show a detailed comparison between the predictions and uncertainties produced by DE and VBNNs trained with different approaches. The MLIPs were trained  using energies only, as well as using both energies and forces, on a dataset comprising 7,815 structures representing various phases of titanium dioxide (TiO\textsubscript{2})~\citep{Artrith2016} and on the energies coming from QM7 dataset, comprising the atomization energies of 7,165 small organic molecules with up to seven heavy atoms (C, N, O, S)~\citep{blum,rupp2012fast}. To assess how variations in data representation affect predictive accuracy and uncertainty estimation of the different models, we performed the trainings for energies and forces on both the full dataset and a reduced subset. Additional energy-only TiO$_2$ results, including a fourth Bayesian variant based on the Flipout estimator that is incompatible with the joint energy-force likelihood, are also provided for comparison in the Supplementary Information.
Sec.~\ref{sec:Theory} includes the theoretical background of the different techniques used for uncertainty quantification, i.e. DE and VBNN. Sec.~\ref{sec:Framework} outlines the computational framework developed for the present study, describing the dataset, the networks architectures and hyperparameters, and introducing the evaluation metrics used to assess both predictive accuracy and the quality of uncertainty quantification. Finally, the training results and the detailed comparison between the different approaches are reported in Sec.~\ref{sec:Results}.

\section{Uncertainty quantification in neural networks}\label{sec:Theory}

\subsection{Deep Ensembles}

Deep Ensembles (DE) or neural networks ensembles were originally proposed by \citet{Lakshminarayanan2016} as a practical approach to uncertainty quantification in neural networks. In the broadest sense, the term ensemble encompasses a variety of strategies: multihead architectures in which several output heads share a common feature extractor; and bootstrap- or bagging-based approaches, in which individual models are trained on
different subsets of the data.  

In this work, we follow the formulation of \citet{Lakshminarayanan2016}, in which $M$ networks are independently trained on the same dataset with different random initializations. We note that the term deep ensemble is used in the conventional sense of the method, and does not refer to the depth of the individual models. In our case, the networks are rather shallow (two hidden layers of 15 units each), consistent with the Behler--Parrinello architecture adopted in the \ae net framework.

Given an input atomic configuration $x$, each model predicts an energy
$E_i = f_{\theta_i}(x)$. The ensemble mean prediction is given by:

\begin{equation}
    \bar{E} = \frac{1}{M} \sum_{i=1}^{M} E_i,
\end{equation}

and the predictive uncertainty is quantified by the sample standard deviation of
the ensemble predictions:

\begin{equation}
    \sigma_E = \sqrt{ \frac{1}{M - 1} \sum_{i=1}^{M} (E_i - \bar{E})^2 }.
\end{equation}

While DEs are attractive because of their ease of implementation, their main drawback is the absence of a rigorous probabilistic foundation: the ensemble variance provides a practical but heuristic measure of uncertainty, rather than one directly derived from a well-defined Bayesian framework. As a result, while DEs often yield reliable estimates in practice, their theoretical justification remains limited compared to fully Bayesian approaches~\citep{Jospin2020,Loaiza-Ganem2025}.

\subsection{Variational Bayesian Neural Networks}

Unlike DE, Bayesian neural networks explicity model uncertainty in their architecture, by treating weights as random variables with a prior distribution $p(\mathbf{w})$. Given a collection of data $\mathcal{D}$, the goal is to compute the posterior $p(\mathbf{w}|\mathcal{D})$, which captures plausible weight configurations conditioned on observations. This yields a posterior predictive distribution that marginalizes over all possible weight configurations. However, exact Bayesian inference in deep neural networks is computationally intractable, motivating the use of approximate inference techniques.

Variational Inference~\citep{Blei2017} approximates the true posterior $p(\mathbf{w}|\mathcal{D})$ over the weights $\mathbf{w}$, given a dataset $\mathcal{D} = \{(x_i, y_i)\}_{i=1}^N$, with a variational distribution (also known as guide) $q_{\theta}(\mathbf{w})$ parametrized by $\theta$. The values of $\theta$ are then learned such that the variational distribution $q_{\theta}(\mathbf{w})$ is as close as possible to the true posterior $p(\mathbf{w}|\mathcal{D})$~\citep{Jospin2020}. The goal is then to minimize the Kullback-Leibler (KL) divergence~\citep{Kullback1951}, which is a measures of similarity of two distributions, between the variational approximation and the true posterior:

\begin{equation}
    D_{KL}(q_{\theta}(\mathbf{w})\,\|\,p(\mathbf{w}|\mathcal{D})) = \int_\mathbf{w} q_{\theta}(\mathbf{w})\,\log \left(\dfrac{q_{\theta}(\mathbf{w})}{p(\mathbf{w}|\mathcal{D})}\right)d\mathbf{w}
\end{equation}

Nonetheless, in order to obtain $D_{KL}$ one still needs to compute $p(\mathbf{w}|\mathcal{D})$. In order to overcome this issue, the \textit{evidence lower bound} (ELBO) is used as a loss during the training, defined as:

\begin{equation}
    \text{ELBO} = \int_\mathbf{w} q_{\theta}(\mathbf{w})\,\log \left(\dfrac{p(\mathbf{w},\mathcal{D})}{q_{\theta}(\mathbf{w})}\right)d\mathbf{w} = \log p(\mathcal{D}) - D_{KL}(q_{\theta}(\mathbf{w})\,\|\,p(\mathbf{w}|\mathcal{D}))
\end{equation}

Because $\log p(\mathcal{D})$ (the log marginal likelihood) is constant with respect to $\theta$, minimizing $D_{KL}$ is equivalent to maximizing the ELBO. The stochastic gradient descent method used to optimize the ELBO is the stochastic variational inference (SVI)~\citep{Hoffman2013} 

To make variational inference tractable in high-dimensional spaces such as those of deep neural networks, a common simplifying assumption to model the variational distribution $q_{\theta}(\mathbf{w})$ is the \textit{mean-field approximation}. This assumes that the variational distribution factorizes over individual weights:

\begin{equation}
q_{\theta}(\mathbf{w}) = \prod_i q_{\theta_i}(w_i)
\end{equation}

This leads to a fully factorized Gaussian posterior where each weight $w_i$ is modeled with its own mean and variance. While this ignores correlations between weights, treating them as independent variables, it greatly reduces computational complexity and is widely used in practice.

 While mean-field approximation offer a relatively simple representation of the variational distribution, optimizing the ELBO with backpropagation remains computationally unfeasible, since the presence of stochastic parameters make standard backpropagation unable to function correctly through internal nodes~\citep{Buntine1994}. 
 
 A practical implementation of SVI to neural networks is \textit{Bayes by Backprop}~\citep{blundell2015}. It implements the mean-field approximation and employs the reparametrization trick~\citep{kingma2022autoencodingvariationalbayes}, expressing weights as:
\[
\mathbf{w} = \boldsymbol{\mu} + \boldsymbol{\sigma} \odot \boldsymbol{\epsilon}
\]
where $\boldsymbol{\epsilon} \sim \mathcal{N}(0,\mathbf{I})$. This formulation allows gradients to flow through stochastic nodes, providing lower-variance gradient estimates than naive score-function methods and making variational Bayesian inference feasible for neural networks. 

However, \textit{Bayes by Backprop} makes convergence slow compared to the usual gradient descent as the ELBO is evaluated via Monte Carlo sampling and typically a small number of samples are used, often just one, making its estimate noisy~\citep{Jospin2020}. For this reason, it is typically combined with techniques to reduce the gradient variance, tailored to Bayesian neural networks, such  as \textit{local reparametrization trick} (LRT)~\citep{kingma2015lrt} and \textit{Flipout} (FO)~\citep{wen2018flipout}. LRT samples the pre-activations of each data point individually, rather than using a single weight matrix across the batch. This is particularly effective for factorized Gaussian posteriors over the weights in layers performing linear mappings, such as dense or convolutional layers. FO, on the other hand, introduces pseudo-independent perturbations by sampling a rank-one sign matrix per data point. This enables computationally efficient per-example weight sampling while preserving the unbiasedness of gradient estimates, further reducing gradient variance without sacrificing performance.

Finally, variational inference can be treated also outside of the mean-field approximation. Radial BNN (RAD)~\citep{farquhar2020radial} introduces radial transformations that induce global correlations among weights, modeled as:
\[
\mathbf{w} = \boldsymbol{\mu} + \boldsymbol{\sigma} \odot \frac{\boldsymbol{\epsilon}}{\vert\boldsymbol{\epsilon}\vert} \cdot r
\]
where $\boldsymbol{\epsilon} \sim \mathcal{N}(0,\mathbf{I})$ and $r \sim \mathcal{N}(0,1)$ is a scalar shared across all weights in a single sample.
These transformations are supposed to better capture heavy-tailed posteriors, reduce gradient variance, and is expected to scale more favourably with larger datasets due to their greater flexibility and expressivity.

In the present work, we extend the BNN framework to jointly predict energies and forces, the latter computed as the negative gradient of the predicted energy with respect to atomic positions. This requires the network to perform two forward passes with identical weight samples: one to compute the energy, and a second (differentiated) pass to obtain forces via automatic differentiation. This constraint is implemented through the \texttt{poutine.trace}/\texttt{poutine.replay} mechanism in Pyro~\citep{bingham2018pyro}, which records weight samples from the
variational guide and replays them exactly for the force computation.

This design, however, is incompatible with the Flipout gradient estimator. Flipout achieves per-example weight diversity by multiplying activations with random sign matrices generated during each forward pass~\citep{wen2018flipout}. When the force pass is replayed from a recorded trace, the underlying latent weight samples are shared correctly, but the sign matrices are not, leading to energy and force predictions computed under inconsistent effective weights. This breaks the theoretical consistency of the joint likelihood. Consequently, Flipout was excluded from the force-training; energy-only FO results on the TiO$_2$ dataset are reported in the Supplementary Information (Section~S1, Tables~S1--S4 and Figures~S1--S3).

\section{Computational Framework}\label{sec:Framework}

\subsection{TiO$_2$ Dataset}

The first dataset employed in this study consists of 7,815 structures of various titanium dioxide (TiO$_2$) phases, originally developed to test the \ae net framework~\citep{Artrith2016}. This dataset, which includes different bulk phases of TiO$_2$, was also used to benchmark the \ae net-PyTorch implementation. The reference energies were obtained from DFT calculations using the Perdew-Burke-Ernzerhof (PBE) exchange–correlation functional.

The dataset was partitioned into training (80\%), validation (10\%), and test (10\%) subsets. The validation set was used for hyperparameter optimization. 

To evaluate the performance and calibration of uncertainty estimates across different data regimes, we conducted experiments using both the full training and validation set (6,330 training and 704 validation structures) and a reduced subset comprising 20\% of the joint training and validation set (1,265 training and 141 validation structures), maintaining the test set constant (781 structures). This setup enabled us to assess and compare model behaviour and uncertainty predictions in both high-data and low-data regimes consistently. This ensures a case in which the training data represent a complete structural picture of the system (high-data regime), and a case in which the system is under-represented (low-data regime). 

\subsection{QM7}

To assess the generalizability of our conclusions beyond a single chemical system,
we additionally evaluate all models on a second dataset, namely the QM7 dataset~\citep{blum, rupp2012fast}, a widely used benchmark in molecular machine learning. QM7  is a subset of the GDB-13 database and comprises 7,165 organic molecules with up to 23 total atoms, including 7 heavy atoms (C, N, O, S), associated to their atomization energies computed at the PBE0 level of theory. Its chemical diversity, spanning a broad range of bonding environments, functional groups, and molecular sizes, makes it a widely used benchmark for assessing the transferability of MLIPs across chemistries.

For this evaluation, the training was performed using the same descriptors used for TiO$_2$ compatible with the \ae net architecture. We adopted the same training protocol as for TiO$_2$: an 80/10/10  train/validation/test split, with hyperparameter optimization performed via Optuna using 60 trials of 3000 epochs each. In contrast to the TiO$_2$ experiments, QM7 structures do not include force labels; models are therefore trained on atomization energies only. In addition, the training was performed only in the high-data regime for this dataset, using 100\% of the data.

\subsection{Model Architectures}

\subsubsection{Neural Networks}

All neural networks used in this study shared the same architecture: two hidden layers with 15 units each and hyperbolic tangent (\texttt{tanh}) activation functions. Atomic environments are represented using Behler--Parrinello atom-centred symmetry functions (ACSFs)~\citep{Behler2007}. For the TiO$_2$ dataset, 70 ACSFs are used per species (Ti and O), yielding 1,321 trainable parameters per element-specific sub-network and 2,642 parameters in total. For QM7, 490 ACSFs per species are required to describe all pairwise and angular interactions across the five elements present (C, H, N, O, S), resulting in 7,621 parameters per sub-network and 38,105 parameters in total. This lightweight configuration provides a balance between expressivity and computational efficiency. The models were implemented using PyTorch, with training managed via the PyTorch Lightning framework~\citep{Falcon_PyTorch_Lightning_2019} to ensure modularity and reproducibility.

\subsubsection{Variational Inference}

For variational inference, we tested two types of variational guides. The first is based on the mean-field approximation, using the AutoNormal guide as implemented in TyXe~\citep{Ritter2021}, which approximates the posterior with a fully factorized diagonal Gaussian. The ELBO is optimized via \textit{Bayes by Backprop}~\citep{blundell2015} with the LRT gradient estimator for variance reduction. Because the posterior is a factorized Gaussian, the KL divergence term in the ELBO admits a closed-form solution; we exploit this by using the \texttt{TraceMeanField\_ELBO} estimator~\citep{bingham2018pyro}, which computes the KL analytically and estimates only the likelihood term via Monte Carlo, further reducing gradient variance compared to a fully stochastic ELBO. The other guide tested is the Radial guide (RAD), which performs a radial transformation over the weights and does not admit an analytic KL; it is therefore optimized with the standard \texttt{Trace\_ELBO} estimator, which estimates the full ELBO via Monte Carlo sampling.

\subsection{Model Training}

All models were trained under both high-data and low-data regimes, for the TiO$_2$ dataset, and under only high-data regime for QM7 datasets. In the high-data regime, each dataset was partitioned into an 80\% training set, a 10\% validation set for hyperparameter optimization, and a 10\% test set for performance evaluation. In the low-data regime, the test set remained fixed, while only 20\% of the entire dataset was used for training and validation (with an 80/20 split within that subset).

The Deep Ensemble (DE) model consisted of 5 independently trained neural networks, each optimized using a weighted root mean squared error (RMSE) loss combining energies and forces:

\begin{equation}
    \mathcal{L} = (1 - \alpha)\,\mathrm{RMSE}_E + \alpha\,\mathrm{RMSE}_F,
\end{equation}

where $\alpha = 0.1$ was chosen from the reference~\cite{Artrith2016} and kept fixed across all models and data regimes to ensure a fair comparison.

In contrast, the Bayesian Neural Networks (BNNs) were trained using variational inference by maximizing the evidence lower bound (ELBO).

To ensure robustness and mitigate the influence of random initialization, each model configuration was trained five times using different random seeds. All models were trained for a maximum of 50,000 epochs with early stopping based on validation loss. A patience of 1500 epochs was applied for both DE and BNNs to prevent overfitting and ensure convergence.

For the TiO$_2$ dataset, we consider models with and without including forces in the training, leading to energy-forces and energy-only models, respectively. The main discussion focuses on the energy-forces model, and the corresponding energy-only results (including the additional FO model variant) are reported in the Supplementary Information (Tables~S1--S4 and Figures~S1--S3). 

Forces are obtained as the negative gradient of the predicted
energy with respect to atomic positions:
\begin{equation}
    \mathbf{F}_i = -\frac{\partial E}{\partial \mathbf{r}_i},
\end{equation}
where $\mathbf{r}_i$ is the position of atom $i$. For the BNN models, forces are incorporated directly into the ELBO via a joint energy-force likelihood:
\begin{equation}
    \log p(\mathbf{E}, \mathbf{F} \mid \mathbf{w}, \mathbf{X}) =
    \log \mathcal{N}(\mathbf{E}; \hat{\mathbf{E}}(\mathbf{w}, \mathbf{X}), \sigma_E^2) +
    \log \mathcal{N}(\mathbf{F}; \hat{\mathbf{F}}(\mathbf{w}, \mathbf{X}), \sigma_F^2),
\end{equation}

where $\hat{\mathbf{E}}(\mathbf{w}, \mathbf{X})$ and $\hat{\mathbf{F}}(\mathbf{w}, \mathbf{X})$ are the predicted energies and forces under weight sample $\mathbf{w}$, and $\sigma_E^2$, $\sigma_F^2$ are the energy and force observation variances fixed as hyperparameters and tuned via Optuna. Crucially, both predictions are evaluated under the same weight sample, ensuring theoretical consistency of the joint likelihood. 
At prediction time, energy mean and uncertainty are both estimated from $N_s = 20$ weight samples drawn from the variational posterior. On the other hand, the force mean is computed at the posterior mean weights (a single deterministic forward pass) while the force uncertainty is instead estimated again from $N_s = 20$ weight samples drawn from the variational posterior. This gives more stable predictions than MC averaging, as the wide BNN posterior causes individual weight samples to produce highly variable force estimates that degrade when averaged.

\subsection{Evaluation Metrics}
\label{sec:metrics}
To assess the accuracy and reliability of the predictive models, we employ several complementary evaluation metrics that quantify both point prediction performance and the quality of uncertainty quantification (UQ).

For model accuracy, we report the Mean Absolute Error (MAE), the Root Mean Squared Error (RMSE), and the Negative Log-Likelihood (NLL). These metrics are meant to provide different insights into the prediction error: MAE captures the average deviation regardless of direction, RMSE emphasizes larger errors and is more sensitive to outliers, and NLL offers a combined score that evaluates both the accuracy and the quality of the uncertainty estimates~\citep{Tran2020}. The MAE, RMSE and NLL are defined as:

\begin{equation}
\mathrm{MAE} = \frac{1}{N} \sum_{i=1}^{N} \left| \hat{y}_i - y_i \right|,
\end{equation}
\begin{equation}
\mathrm{RMSE} = \sqrt{\frac{1}{N} \sum_{i=1}^{N} (\hat{y}_i - y_i)^2},
\end{equation}
\begin{equation}
\mathrm{NLL} = \frac{1}{N} \sum_{i=1}^{N} \left[ \frac{(y_i - \hat{y}_i)^2}{2\sigma_i^2} + \frac{1}{2} \log(2\pi\sigma_i^2) \right],
\end{equation}

where $\hat{y}_i$ and $y_i$ denote the predicted and true values, respectively, and $\sigma_i^2$ is the predictive variance for data point $i$.

To evaluate the quality of uncertainty quantification (UQ), we employ calibration metrics, which assess how well predicted confidence intervals correspond to actual outcome frequencies. In particular, we use calibration curves, which plot the empirical frequency of true values falling within a given predicted confidence level. Ideally, a model's 90\% confidence intervals should contain the true value 90\% of the time. Calibration curves thus provide a direct visual assessment of how well uncertainty estimates match empirical behavior~\citep{pmlr-v80-kuleshov18a}.

We also quantify miscalibration using the Root Mean Squared Calibration Error (RMSCE), which measures the area between the calibration curve and the diagonal representing perfect calibration. Lower RMSCE values indicate better uncertainty estimates.

\begin{figure}[h]
\centering
\includegraphics[width=0.95\linewidth]{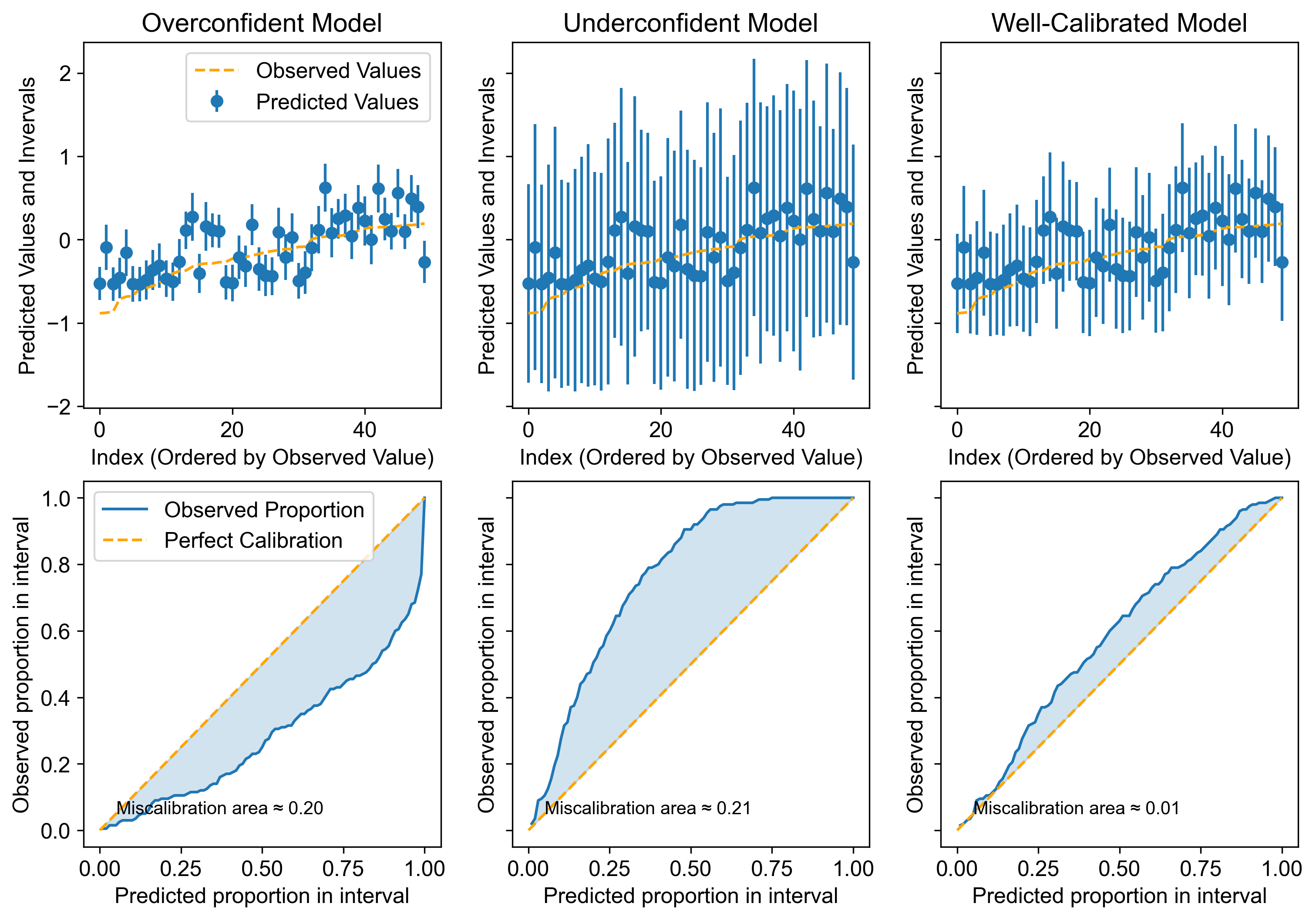}
\caption[Calibration curves examples]{Illustration of uncertainty calibration behavior in three types of models. Top: predicted values ordered by observed values. Bottom: calibration curves showing observed vs. predicted proportions of values within confidence intervals. Left: overconfident model; center: underconfident model; right: well-calibrated model.}
\label{fig:calibration_plots}
\end{figure}

Figure~\ref{fig:calibration_plots} illustrates three prototypical cases of model calibration:

\begin{itemize}
    \item Left: An overconfident model predicts uncertainty intervals that are too narrow, failing to encompass the true values as often as expected. This results in a calibration curve that lies below the diagonal and a substantial miscalibration area.
    \item Center: An underconfident model produces overly broad intervals, capturing the true values more frequently than required. The corresponding calibration curve lies above the diagonal, reflecting an overly cautious model.
    \item Right: A well-calibrated model yields prediction intervals that closely match empirical coverage across all confidence levels. Its calibration curve follows the diagonal closely, and the miscalibration area is minimal.
\end{itemize}

These examples underscore the importance of both accurate and well-calibrated uncertainty estimates, particularly when predictive models are applied to tasks where trust and interpretability are critical.

However, calibration alone does not fully characterize the informativeness of uncertainty estimates. A model can be well-calibrated while still producing overly broad predictions. To complement calibration, we therefore compute the Sharpness, which measures the average predictive uncertainty:

\begin{equation}
\mathrm{Sharpness} = \frac{1}{N} \sum_{i=1}^{N} \sigma_i.
\end{equation}

Here, $\sigma_i$ is the predicted standard deviation for data point $i$. Lower sharpness indicates more confident (narrower) predictions. In an ideal model, low sharpness is achieved without compromising calibration, yielding both informative and trustworthy uncertainty estimates.

In the context of a hypothetical active learning scenario, we introduce two additional metrics to assess the alignment between predicted uncertainties and actual prediction errors. The first is the Spearman rank correlation coefficient ($r_s$) between the predicted uncertainty (standard deviation $\sigma$) and the absolute prediction error. $r_s$ captures monotonic relationships without assuming linearity or Gaussian distributions, making it well suited for comparing uncertainty and error distributions that may be heavy-tailed. It is defined as:

\begin{equation}
    r_s = 1 - \frac{6 \displaystyle\sum_{i=1}^{n} d_i^2}{n(n^2 - 1)},
\end{equation}

where $d_i = \mathrm{rank}(\sigma_i) - \mathrm{rank}(e_i)$ is the difference between the rank of the predicted uncertainty and the rank of the absolute error for data point $i$, and $n$ is the total number of samples. A higher $r_s$ indicates a stronger monotonic correlation between predicted uncertainty and actual prediction error, which is desirable in active learning where uncertainty is used to guide data acquisition.

The second metric, which we name the \textbf{Overlap Score}, is designed to capture the consistency between high-error and high-uncertainty regions-areas of particular interest for active learning algorithms, which typically prioritize samples where model confidence is low and prediction error is high. By quantifying the alignment between these regions, the overlap score provides a practical diagnostic for evaluating the suitability of uncertainty estimates in guiding data acquisition. Both the predicted uncertainties and the corresponding absolute errors are discretized into quartiles. The Overlap Score is defined as the percentage of high-uncertainty predictions that also lie in the high-error region. This answers the question: Among the data points the model is least certain about, how many are actually among the most inaccurate? This makes the metric particularly suitable for evaluating the usefulness of uncertainty estimates in active learning, quantifying how many of the points one would use to perform new calculations and re-train the model were actually useful.

Formally, let the true target values be \( \mathbf{y}_{\text{true}} = \{ y_1^{\text{true}}, \dots, y_n^{\text{true}} \} \), the predicted values \( \mathbf{y}_{\text{pred}} = \{ y_1^{\text{pred}}, \dots, y_n^{\text{pred}} \} \), and the predicted standard deviations (uncertainties) \( \boldsymbol{\sigma} = \{ \sigma_1, \dots, \sigma_n \} \). Define the absolute error for each sample as

\begin{equation}
    e_i = \left| y_i^{\text{true}} - y_i^{\text{pred}} \right|.
\end{equation}

Let \( Q_3^{\text{error}} \) and \( Q_3^{\text{uncertainty}} \) be the third quartile (75th percentile) of the absolute error and uncertainty distributions, respectively. Then, define the indicator functions

\[
H_i^{\text{error}} = 
\begin{cases}
1 & \text{if } e_i > Q_3^{\text{error}} \\
0 & \text{otherwise}
\end{cases}
\quad\text{and}\quad
H_i^{\text{uncertainty}} = 
\begin{cases}
1 & \text{if } \sigma_i > Q_3^{\text{uncertainty}} \\
0 & \text{otherwise}.
\end{cases}
\]

The overlap score is then given by

\begin{equation}
    \text{Overlap Score} = 100 \times \frac{ \sum_{i=1}^n H_i^{\text{error}} \cdot H_i^{\text{uncertainty}} }{ \sum_{i=1}^n H_i^{\text{uncertainty}} }.
\end{equation}

\begin{figure}[h]
    \centering
    \includegraphics[width=0.7\textwidth]{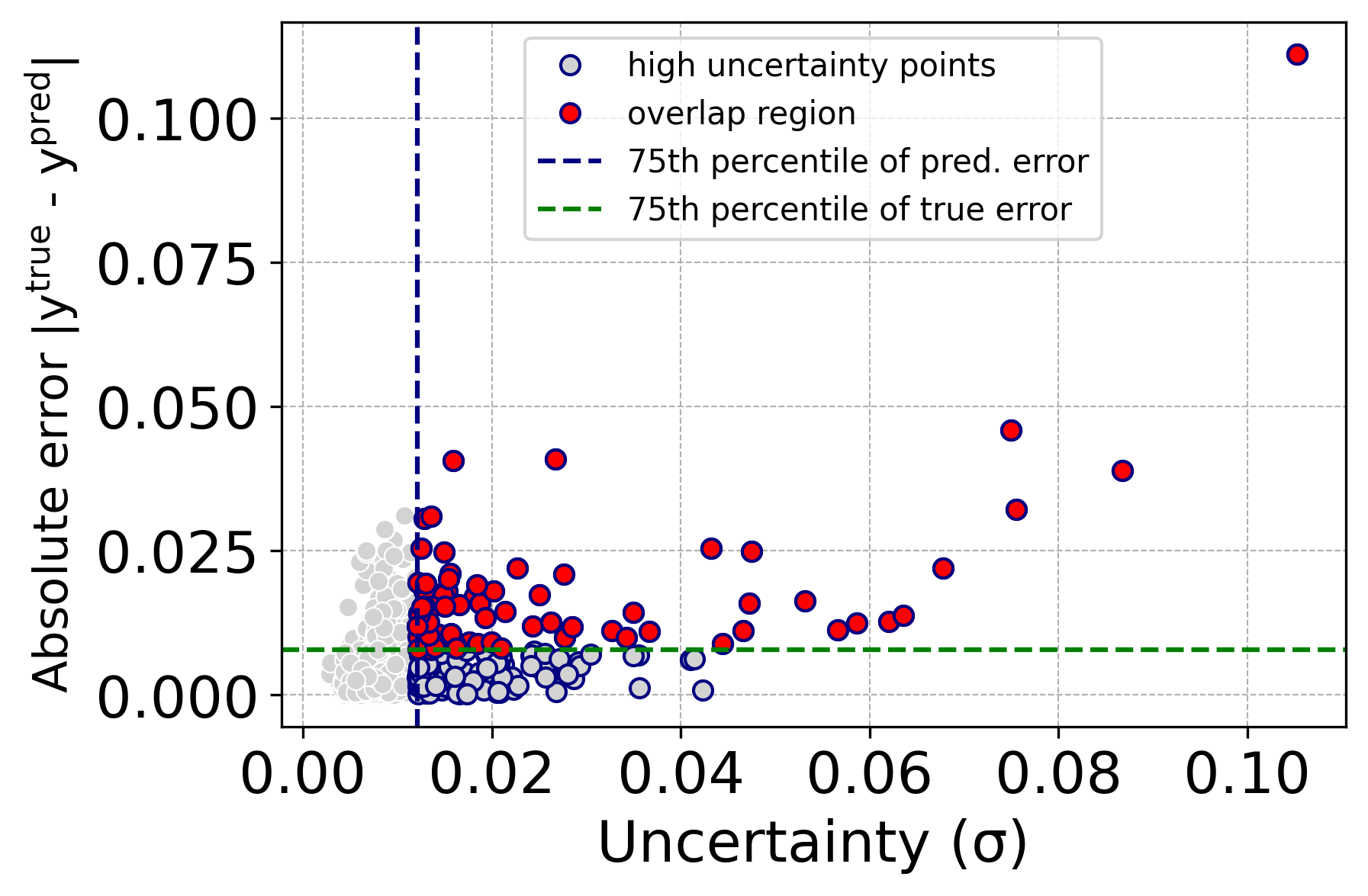}
    \caption[Overlap metric example]{Illustration of the overlap metric. Each point represents a prediction with its uncertainty (x-axis) and absolute error (y-axis). Points outlined in blue correspond to cases with uncertainties above the third quartile (high-uncertainty region). Point highlighted in red represent high-uncertainty, high true error cases in the upper right part of the plot. The overlap metric is defined as the ratio between the number of red points and the total number of blue outlined points, quantifying how often high-uncertainty predictions correctly identify high-error cases.}
    \label{fig:overlap_example}
\end{figure}

An illustration of the overlap metric is provided in Figure~\ref{fig:overlap_example}, where each point represents a single prediction plotted by its associated uncertainty (x-axis) and absolute error (y-axis). The red points in the upper-right quadrant highlight cases of simultaneously high uncertainty and high error, whose density reflects the degree of overlap.

\subsection{Hyperparameter Optimization}

The optimal hyperparameters selected via Optuna for the TiO$_2$ low-data regime, TiO$_2$ high-data regime and QM7 trainings are reported in Tables~\ref{tab:optuna_hyperparams_20},~\ref{tab:optuna_hyperparams_100},
and~\ref{tab:optuna_hyperparams_qm7}, respectively.

The number of MC samples during training varies between 1 and 2 depending on the
model and regime. LRT generally favours a single MC sample, reflecting its
lower-variance gradient estimates from the local reparametrization. RAD shows more
variability in this choice across datasets.

Learning rates are consistently low ($10^{-5}$--$10^{-4}$) for the Bayesian
models, with slightly higher values in the low-data regime. This is consistent
with the reduced number of gradient steps available before early stopping triggers,
where a more aggressive rate helps convergence without compromising generalisation.
DE selects substantially higher learning rates ($10^{-3}$--$10^{-2}$), consistent
with the simpler, deterministic training objective.

The prior scale and variational posterior scale ($q_\theta$) indicate differing
regularisation levels. In the high-data regime, models tend to prefer tighter
posteriors (lower $q_\theta$), reflecting the larger training set supporting more
confident posterior distributions. In the low-data regime, $q_\theta$ increases
for most models, retaining more parameter uncertainty due to limited data.

The observation noise scale $\sigma_E$ is generally higher in the low-data regime,
suggesting that models attribute more residual error to aleatoric uncertainty when
data are scarce.

Batch size choices reflect a trade-off between gradient noise and computational
efficiency. LRT favours smaller batches (128--256), while RAD tends to select
larger batches (256--512).

\begin{table}[h]
\centering
\caption[Hyperparameters low-data regime]{Best hyperparameters selected via Optuna
for each model trained on 20\% of the TiO$_2$ dataset.}
\label{tab:optuna_hyperparams_20}
\begin{tabular}{lccc}
\toprule
\textbf{Hyperparameter} & \textbf{LRT} & \textbf{RAD} & \textbf{DE} \\
\midrule
Learning Rate      & $2.42\times10^{-5}$ & $3.57\times10^{-4}$ & $2.73\times10^{-3}$ \\
Batch Size         & 256   & 512   & 128 \\
MC Samples         & 1     & 2     & -- \\
Prior Scale        & 0.354 & 0.156 & -- \\
$q_\theta$ Scale   & $1.27\times10^{-3}$ & $2.26\times10^{-3}$ & -- \\
$\sigma_E$         & 0.226 & 1.026 & -- \\
$\sigma_F$         & 0.139 & 0.888 & -- \\
\bottomrule
\end{tabular}
\end{table}


\begin{table}[h]
\centering
\caption[Hyperparameters high-data regime]{Best hyperparameters selected via Optuna
for each model trained on 100\% of the TiO$_2$ dataset.}
\label{tab:optuna_hyperparams_100}
\begin{tabular}{lccc}
\toprule
\textbf{Hyperparameter} & \textbf{LRT} & \textbf{RAD} & \textbf{DE} \\
\midrule
Learning Rate      & $2.75\times10^{-5}$ & $4.23\times10^{-5}$ & $1.02\times10^{-3}$ \\
Batch Size         & 128   & 256   & 256 \\
MC Samples         & 1     & 1     & -- \\
Prior Scale        & 0.392 & 0.348 & -- \\
$q_\theta$ Scale   & $2.03\times10^{-5}$ & $6.04\times10^{-5}$ & -- \\
$\sigma_E$         & 0.191 & 0.180 & -- \\
$\sigma_F$         & 0.386 & 0.140 & -- \\
\bottomrule
\end{tabular}
\end{table}


\begin{table}[h]
\centering
\caption[Hyperparameters QM7]{Best hyperparameters selected via Optuna for each
model trained on the QM7 dataset (energy-only; no force training).}
\label{tab:optuna_hyperparams_qm7}

\begin{tabular}{lccc}
\toprule
\textbf{Hyperparameter} & \textbf{LRT} & \textbf{RAD} & \textbf{DE} \\
\midrule
Learning Rate      & $4.65\times10^{-5}$ & $6.85\times10^{-5}$ & $8.49\times10^{-3}$ \\
Batch Size         & 256   & 256   & 128 \\
MC Samples         & 1     & 2     & -- \\
Prior Scale        & 0.169 & 0.191 & -- \\
$q_\theta$ Scale   & $2.49\times10^{-5}$ & $2.36\times10^{-4}$ & -- \\
$\sigma_E$         & 0.113 & 0.121 & -- \\
\bottomrule
\end{tabular}
\end{table}

\section{Results and Discussion}\label{sec:Results}
\subsection{Predictive Accuracy}
\label{sec:results_performance}

\begin{figure}[h]
\centering
\includegraphics[width=\textwidth]{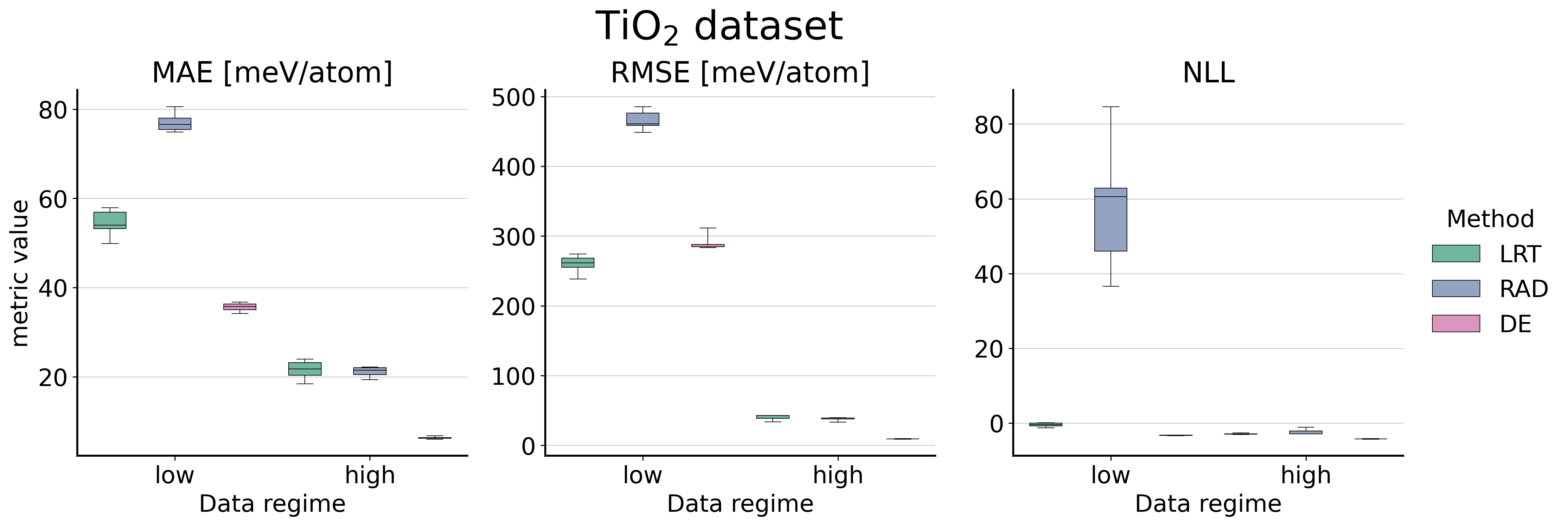}
\caption[Mean absolute error, root mean square error, negative log likelihood]{Predictive performance metrics (MAE, RMSE, NLL) for all models trained on TiO$_2$ across high-data (100\% of the dataset) and low-data (20\% of the dataset) regimes. Boxplots summarize the distribution of scores over five independent training runs: the boxes show the interquartile range (25th to 75th percentile), the horizontal line within each box marks the median, and the whiskers extend to the whole range of the scores. Colour coding: LRT (green), RAD (orange) and DE (blue). The associated values are reported in table~\ref{tab:avg_metrics_tio2}.}
\label{fig:performance_de_bnn}
\end{figure}

The predictive accuracy of each model is summarized in Figure~\ref{fig:performance_de_bnn}, which reports the results for both the high-data and low-data regimes. The evaluation metrics considered are MAE, RMSE, and NLL, as introduced in Section~\ref{sec:metrics}. Each box in the figure represents the distribution of the metric values obtained from five independently trained models using different random seeds. This allows us to assess the robustness and variability of each method under repeated training.

Table~\ref{tab:avg_metrics_tio2} reports the predictive performance of each method for both data regimes for TiO$_2$ dataset. 

All models, when trained with the whole dataset, achieved relatively low MAE and RMSE values, confirming the effectiveness of both ensemble and variational strategies when sufficient data are available.

DE achieved the best performance across all metrics, with MAE of 6 meV/atom and RMSE of 9, outperforming the Bayesian counterparts. LRT and RAD followed closely, both showing similar MAE values of 21 meV/atom, and RMSE of 39 and 38 meV/atom, respectively. For reference, the same models trained on TiO$_2$ without forces (Table~S3) reach lower energy MAE (e.g.\ LRT 14 vs.\ 21 meV/atom), reflecting the additional training complexity introduced by the joint energy-force likelihood.

The negative log-likelihood (NLL) values further underscore the better performance of DE. Indeed, DE obtained the most negative NLL at -4.2, indicating not only high predictive accuracy but also well-calibrated uncertainty estimates. LRT (-2.9) and RAD (-2.2) also provided reasonable calibration, but worse performance overall.

\begin{table}[h]
\centering
\caption[Accuracy performances of high-data regime]{Average MAE, RMSE, and NLL for each model trained on 20\% and 100\% of the dataset. All values are reported in meV/atom, with uncertainty indicating the standard deviation over five independent runs.}
\label{tab:avg_metrics_tio2}
\begin{tabular}{llccc}
\toprule
\textbf{Regime} & \textbf{Method} & \textbf{MAE [meV/atom]} & \textbf{RMSE [meV/atom]} & \textbf{NLL} \\
\midrule
\textbf{High-data}
& LRT &  21 ± 2 & 39 ± 4 & -2.9 ± 0.2 \\
& RAD  & 21 ± 1 & 38 ± 3 & -2.2 ± 0.7  \\
& DE & 6 ± 0.2 & 9 ± 0.3 & -4.2 ± 0.1 \\
\midrule
\textbf{Low-data}
& LRT     & 54 ± 3 & 259 ± 14 & -0.5 ± 0.5 \\
& RAD     & 77 ± 2 & 466 ± 15 & 58 ± 18  \\
& DE      & 35 ± 1 & 291 ± 11 & -3.3 ± 0.1 \\
\bottomrule
\end{tabular}
\end{table}

In the low-data regime, despite the reduced training set, DE retained its overall superiority in accuracy, achieving the lowest MAE of 35 meV/atom and RMSE of 291. Among the Bayesian models, LRT remained competitive, achieving a similar accuracy with a MAE of 54 and even a lower RMSE of 259 when compared to DE, suggesting a higher robustness to outliers. Nonetheless, DE kept a better performance in terms of NLL respect to LRT, showing a better calibration of errors even for smaller datasets.

RAD, on the other hand, offered the lowest accuracy among the three models and a particularly poor performance in terms of NLL. This is due to the presence of big outliers in the predictions, as shown by the highest value of the RMSE among the three, coupled with very low predicted uncertainties, which suggest that such scheme suffers under-representative dataset.

In summary, DE consistently achieved the lowest values across all evaluated metrics-including MAE, RMSE, and especially NLL, highlighting its superior predictive accuracy and more reliable uncertainty quantification compared to the Bayesian alternatives.

Among the Bayesian approaches, LRT yielded comparable results and, notably, offered a better performance in terms of RMSE in the low-data regime. This reduced sensitivity to extreme prediction errors suggests that variational strategies such as LRT may offer enhanced generalization capabilities under low-data conditions.

Conversely, RAD performed the worst in the low-data regime, while giving similar performance to LRT in the high-data one. Its elevated error metrics when using a limited dataset and substantially worse NLL values suggest instability during training, likely due to the incompatibility of the RAD guide with low-variance gradient estimators or due to the limited size of the model.

It should also be noted that Bayesian approaches appear to be more influenced by the initial random initialization compared to DE. This is most likely due to the higher complexity of the training and to the noisy gradients.

\subsection{Uncertainty Quantification}
\label{sec:results_uncertainty}

\begin{figure}[h]
\centering
\includegraphics[width=0.8\textwidth]{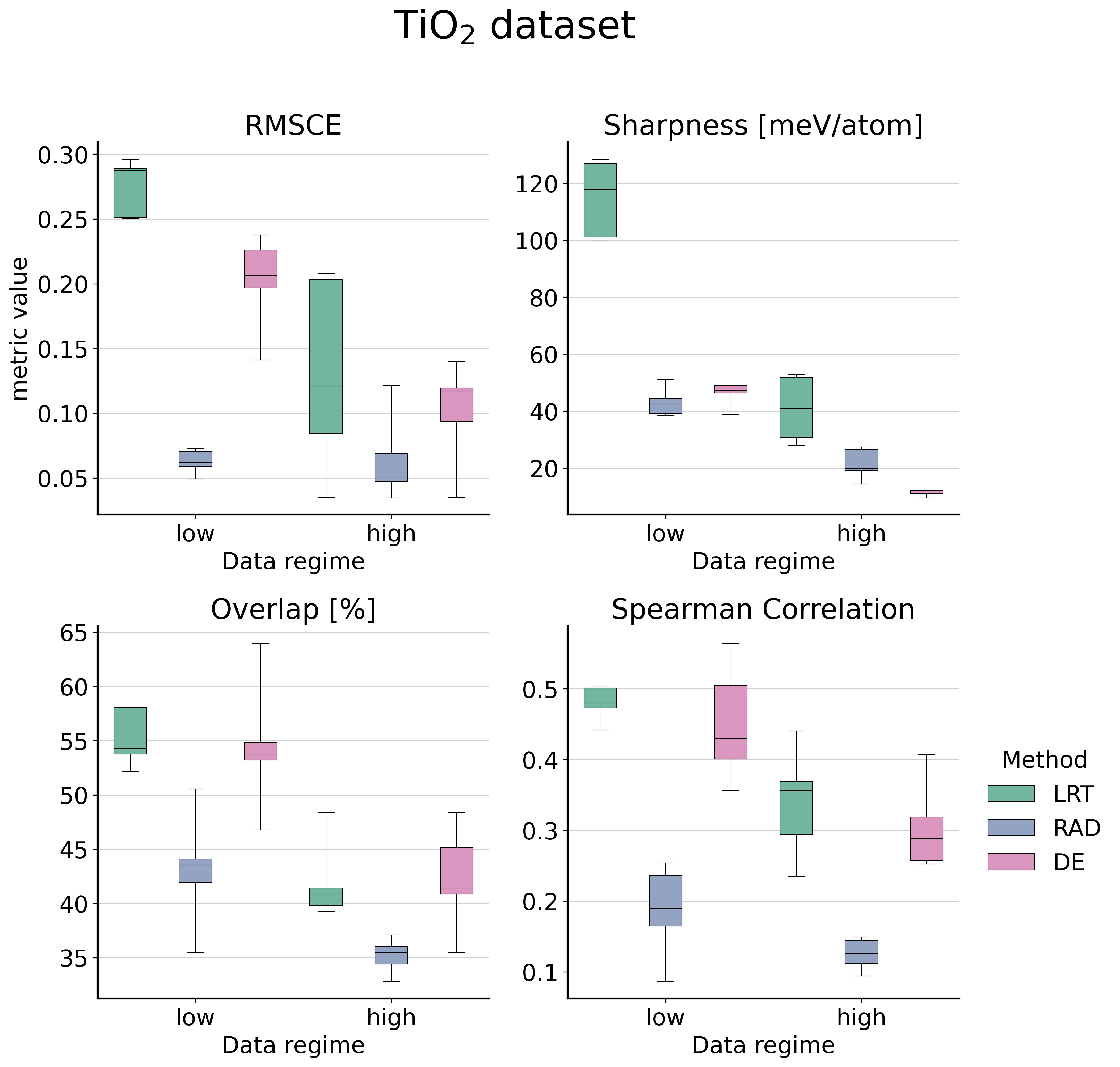}
\caption[Root mean squared calibration error and sharpness]{Root mean squared calibration error (RMSCE), sharpness, overlap and Spearman correlation (r$_s$) of the predicted uncertainty distributions across methods and data regimes for both the data regime of the TiO$_2$. Box plots summarize the distribution of scores over five independent training runs: the boxes show the interquartile range (25th to 75th percentile), the horizontal line within each box marks the median, and the whiskers extend to the most extreme points. Colour coding: LRT (green), RAD (orange) and DE (blue). The associated values are reported in table~\ref{tab:uq_metrics_tio2}.}
\label{fig:uq_quality}
\end{figure}

The uncertainty quantification (UQ) performance of each model is summarized in Figure~\ref{fig:uq_quality}, which presents results for both the high-data and low-data regimes. The evaluation relies on four metrics: root mean squared calibration error (RMSCE), sharpness, overlap score and Spearman rank correlation coefficient ($r_s$), as defined in Section~\ref{sec:metrics}.

Table~\ref{tab:uq_metrics_tio2} shows the metrics mean and standard deviation of the three models. In high-data regime, all the models performed well in terms of calibration. RAD achieved the best overall calibration with the lowest RMSCE of 0.06, highlighting its ability to produce well-calibrated uncertainty estimates when sufficient training data are available. LRT and DE follow it with slightly higher calibration errors.

In terms of sharpness, DE again outperformed the alternatives, exhibiting the lowest value (11 meV/atom), indicative of confident and concentrated predictive distributions. RAD  produced the best sharpness results between the two bayesian models (21). LRT, on the other hand, with the highest sharpness value (41), produced broader uncertainty intervals, consistent with a tendency to overestimate predictive uncertainty.

\begin{table}[h]
\centering
\caption[Uncertainty quantification of high-data regime]{Average RMSCE, Sharpness, Overlap and Spearman correlation (r$_s$) for each model trained on 20\% and 100\% of the dataset. All values are reported in eV/atom, with uncertainty indicating the standard deviation over five independent runs.}
\label{tab:uq_metrics_tio2}
\begin{tabular}{llcccc}
\toprule
\textbf{Regime} & \textbf{Method} 
& \textbf{RMSCE} 
& \textbf{Sharpness [meV/atom]} 
& \textbf{Overlap [\%]} 
& \textbf{r$_s$} \\
\midrule

\textbf{High-data}

& LRT & $0.13 \pm 0.08$ & $41 \pm 12$ & $42 \pm 4$ & $0.34 \pm 0.08$ \\
& RAD & $0.06 \pm 0.03$ & $21 \pm 5$ & $35 \pm 2$ & $0.13 \pm 0.02$ \\
& DE  & $0.10 \pm 0.04$ & $11 \pm 1$ & $42 \pm 5$ & $0.30 \pm 0.06$ \\
\midrule

\textbf{Low-data}
& LRT & $0.27 \pm 0.02$ & $115 \pm 14$ & $55 \pm 3$ & $0.28 \pm 0.02$ \\
& RAD & $0.05 \pm 0.01$ & $45 \pm 4$ & $38 \pm 5$ & $0.19 \pm 0.07$ \\
& DE  & $0.20 \pm 0.04$ & $46 \pm 4$ & $55 \pm 6$ & $0.45 \pm 0.08$ \\
\bottomrule
\end{tabular}

\end{table}

While handling the smaller and less complete dataset of the low-data regime, the models exhibited a degradation in calibration quality, as indicated by increased RMSCE values, except for RAD that remarkably kept a similar calibration error to the one obtained in the high-data regime. It achieved the best calibration (RMSCE = 0.05) and the sharpest uncertainty estimates (45) together with DE, suggesting it offered the most favourable UQ performance in this setting. DE ranked second in calibration (RMSCE = 0.20) with a sharpness value similar to RAD. Both LRT and DE roughly doubled their RMSCE relative to the high-data regime (0.13 $\rightarrow$ 0.27 and 0.10 $\rightarrow$ 0.20, respectively), but LRT showed the largest absolute drop in calibration reliability. Moreover, its very high sharpness value reflects broader and less confident uncertainty estimates.

\begin{figure}[h]
    \centering
    \includegraphics[width=0.8\linewidth]{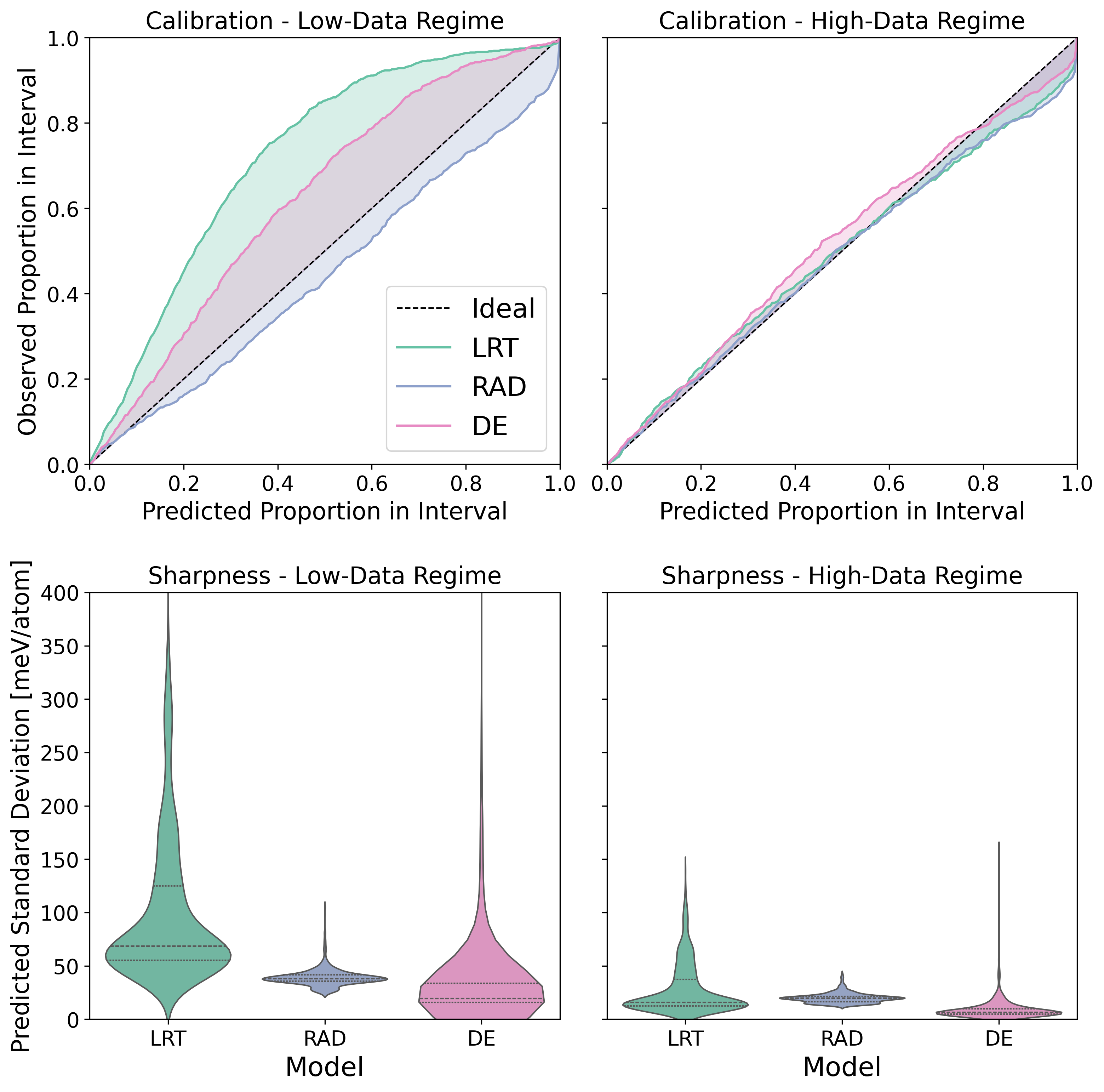}
    \caption[Calibration and sharpness curves for best models]{Calibration curves (top) and sharpness distributions (bottom) for the best-performing models in each data regime, selected based on the lowest RMSCE value.} 
    \label{fig:cal_sharp_best}
\end{figure}

Figure~\ref{fig:cal_sharp_best} provides a visual assessment of the uncertainty quality for the best-performing models in the low-data and high-data regimes, as selected based on the RMSCE metric. In the top row, calibration curves show how closely the predicted confidence intervals match the empirical coverage. A well-calibrated model should align closely with the diagonal dashed line (Ideal). 

Under the low-data regime (left), the RAD model demonstrates a very good calibration overall, showing that Table~\ref{tab:uq_metrics_tio2} does not fully capture the trade-offs between calibration and confidence. This highlights the importance of jointly considering both metrics when evaluating uncertainty quality. In the high-data regime (right), all the models achieve near-ideal calibration.

The lower row of Figure~\ref{fig:cal_sharp_best} depicts sharpness through violin plots of the predicted standard deviations. Narrower and more concentrated distributions near zero indicate more confident (i.e., sharper) predictions. As expected, all models demonstrate improved sharpness in the high-data regime relative to the low-data regime, with DE producing the sharpest distributions in both cases. RAD closely matches DE in calibration performance. In contrast, LRT displays good calibration but a higher sharpness value due to a broader standard deviation distribution.

We further assessed the quality of uncertainty quantification using the Spearman rank correlation ($r_s$) and the overlap score, as shown in Figure~\ref{fig:uq_quality} and reported in the fifth and sixth columns of Table~\ref{tab:uq_metrics_tio2}. In the high-data regime, LRT achieved the highest $r_s$ (0.34), followed closely by DE (0.30), while RAD showed a notably weaker correlation (0.13). In the low-data regime, DE achieved the highest $r_s$ (0.45), with LRT and RAD showing reduced values of 0.28 and 0.19, respectively, indicating that under data scarcity DE maintains the strongest monotonic correlation between predicted uncertainty and actual prediction error.

Regarding the overlap score, in the high-data regime LRT and DE achieved comparable scores (42\%), while RAD underperformed (35\%). Under the low-data regime, DE and LRT again tied at the highest overlap (55\%), while RAD showed the weakest performance (38\%). These results indicate that DE and LRT are comparably effective at identifying high-error regions across both data regimes, with RAD consistently underperforming in this regard.

Figures S1, S2 and S3, together with Tables S3 and S4 report the performance of the three NN architectures (LRT, RAD, and DE) when training only on energies (energy-only models). For the energy-only models, we also evaluate the Flipout (FO) approach to BNN, which is incompatible with force-training (see methods section). In the energy-only setting (Tables~S3 and~S4), FO achieved accuracy comparable to LRT, with calibration on par in the low-data regime but degraded at high data (RMSCE 0.19 vs 0.08).

Comparing the joint energy-force training to the energy-only setting
(Tables~S3 and~S4) allows for an  estimate of the cost of incorporating forces into the
objective. In the high-data regime, energy-only training yields lower energy
errors for the BNN models, with the energy MAE of LRT decreasing from
21 to 14~meV/atom ($33\%$ higher accuracy) and that of RAD from 21 to 19~meV/atom, while DE
is essentially unchanged (6 vs.\ 5~meV/atom); calibration follows the same
trend, with the RMSCE of LRT and DE roughly halving (from 0.13 to 0.08
and from 0.10 to 0.05, respectively). The notable exception is RAD,
whose high-data calibration is markedly better with forces
(RMSCE 0.06 vs.\ 0.19), suggesting that the force signal stabilises this
guide. The effect is far more pronounced in the low-data regime, where the
joint likelihood induces large energy outliers: removing forces lowers the
energy MAE by roughly $50$-$67\%$ (LRT 54 to 33, RAD
77 to 46, DE 35 to 23~meV/atom) and the energy RMSE by a factor of $3.5$-$5$ (e.g.\ RAD 466 to 96~meV/atom), while the pathological NLL of RAD (58) is recovered to a sensible value ($-2.15$).
Uncertainty estimates also become sharper and equally or better calibrated for LRT and DE in this setting.

It is worth noting that part of the gap reflects the $\alpha$-weighted energy-force loss balancing two targets. Crucially, the relative ordering of the methods (DE~$>$~LRT~$>$~RAD) is preserved in both settings.

\subsection{Force Predictions and Uncertainty}
\label{sec:results_forces}

Table~\ref{tab:force_metrics} summarizes the predictive performance of the models regarding the forces. The table reports RMSE, MAE, and RMSCE for the three models in both data regimes. All
models trained with force labels achieve a reasonable correlation with the reference
data in the high-data regime. DE achieves the lowest force RMSE
($0.373 \pm 0.040$~eV/\AA), followed by RAD ($0.580 \pm 0.100$~eV/\AA) and LRT
($0.708 \pm 0.059$~eV/\AA), with similar trends for the MAE values.

\begin{table}[h]
\centering
\caption[Force prediction metrics]{Force RMSE, MAE, and RMSCE for each model in
high-data (100\%) and low-data (20\%) regimes, reporting the average value and the standard deviation obtained over five independent runs.}
\label{tab:force_metrics}
\begin{tabular}{llccc}
\toprule
\textbf{Regime} & \textbf{Method} 
& \textbf{F-MAE [eV/\AA]} 
& \textbf{F-RMSE [eV/\AA]} 
& \textbf{F-RMSCE} \\
\midrule

\textbf{High-data}
& LRT & $0.360 \pm 0.010$ & $0.708 \pm 0.059$ & $0.513 \pm 0.001$ \\
& RAD & $0.282 \pm 0.011$ & $0.580 \pm 0.100$ & $0.450 \pm 0.002$ \\
& DE  & $0.187 \pm 0.003$ & $0.373 \pm 0.040$ & $0.256 \pm 0.004$ \\
\midrule

\textbf{Low-data}
& LRT & $0.426 \pm 0.029$ & $1.654 \pm 0.342$ & $0.423 \pm 0.004$ \\
& RAD & $0.465 \pm 0.012$ & $1.426 \pm 0.088$ & $0.541 \pm 0.001$ \\
& DE  & $0.401 \pm 0.009$ & $2.644 \pm 0.164$ & $0.149 \pm 0.009$ \\
\bottomrule
\end{tabular}
\end{table}

Nonetheless, in the low-data regime, both LRT and RAD achieve substantially lower force RMSE than
DE ($1.654$ and $1.426$~eV/\AA\ vs.\ $2.644$~eV/\AA). This mirrors the energy
results, where Bayesian approaches outperform ensembles under data scarcity, likely
due to the regularization provided by the prior.

Force calibration (F-RMSCE) is generally poor across all models, indicating that force
uncertainty estimates require further improvement and that this remains an open
challenge. Only DE exhibits a good calibration under the low-data regime.

\subsection{Generalization: QM7 Dataset}
\label{sec:results_qm7}

\begin{figure}[h]
\centering
\includegraphics[width=\textwidth]{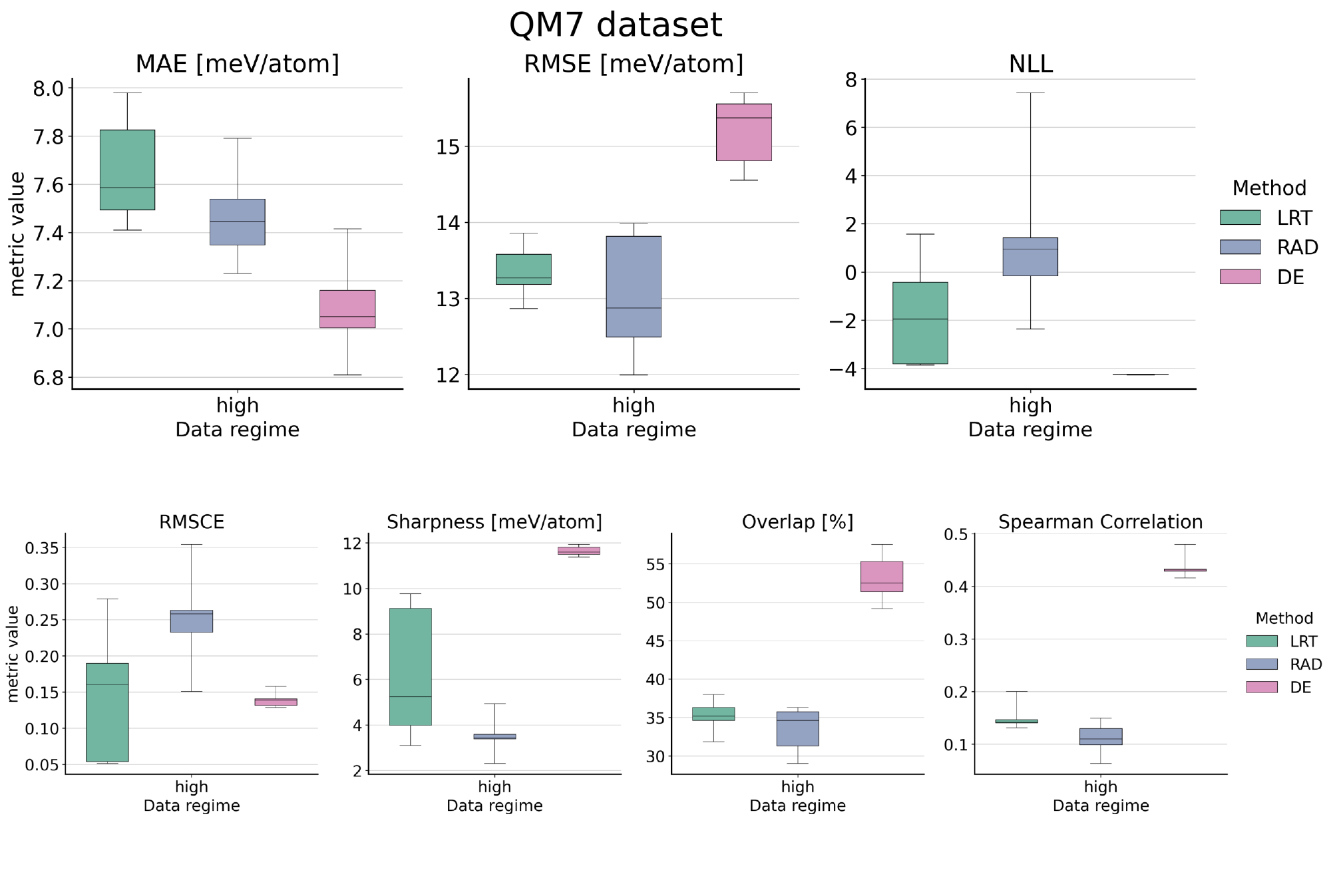}
\caption[Mean absolute error, root mean square error, negative log likelihood]{Predictive performance metrics (MAE, RMSE, NLL) in the top row, together with uncertainty quantification metrics (RMSCE, sharpness, overlap and Spearman correlation) in the bottom one for all models trained on QM7 dataset. Boxplots summarize the distribution of scores over five independent training runs: the boxes show the interquartile range (25th to 75th percentile), the horizontal line within each box marks the median, and the whiskers extend to the whole range of the scores. Colour coding: LRT (green), RAD (orange) and DE (blue). The associated values are reported in tables~\ref{tab:qm7_metrics_en} and ~\ref{tab:qm7_metrics_uq}.}
\label{fig:performance_de_qm7}
\end{figure}

To assess the generalizability of our conclusions to a chemically distinct system,
we evaluate all models on the QM7 benchmark.
Table~\ref{tab:qm7_metrics_en} reports the energy RMSE, MAE, NLL on the
QM7 test set, while the uncertainty related metrics are reported in Table~\ref{tab:qm7_metrics_uq}. All the metrics are summarized in Figure~\ref{fig:performance_de_qm7}.

\begin{table}[h]
\centering
\caption[QM7 performance metrics]{Average accuracy metrics (MAE, RMSE, and NLL) for each model trained on the QM7 dataset together with their uncertainties indicating the standard deviation over five independent runs.}
\label{tab:qm7_metrics_en}
\begin{tabular}{lccc}
\toprule
\textbf{Method} & \textbf{MAE [meV/atom]} & \textbf{RMSE [meV/atom]} & \textbf{NLL}  \\
\midrule
LRT     & $7.7 \pm 0.2$ & $13.3 \pm 0.4$  & $-1.7 \pm 2.3$  \\
RAD     & $7.5 \pm 0.2$ & $13.0 \pm 0.8$  & $1.5 \pm 3.7$   \\
DE      & $7.1 \pm 0.2$ & $15.2 \pm 0.5$  & $-4.3 \pm 0.01$  \\
\bottomrule
\end{tabular}
\end{table}

\begin{table}[h]
\centering
\caption[QM7 performance metrics uq]{Uncertainty quantification metrics (RMSCE, Sharpness, Overlap and r$_s$) for each model trained on the QM7 dataset, expressed as the average value together with the standard deviation obtained over five independent runs.}
\label{tab:qm7_metrics_uq}
\begin{tabular}{lcccc}
\toprule
\textbf{Method} 
& \textbf{RMSCE} 
& \textbf{Sharpness} 
& \textbf{Overlap [\%]} 
& \textbf{r$_s$} \\
\midrule
LRT & $0.15 \pm 0.10$ & $6 \pm 3$ & $35 \pm 2$ & $0.15 \pm 0.03$ \\
RAD & $0.25 \pm 0.07$ & $4 \pm 1$ & $33 \pm 3$ & $0.11 \pm 0.02$ \\
DE  & $0.14 \pm 0.01$ & $12 \pm 0$ & $53 \pm 3$ & $0.44 \pm 0.02$ \\
\bottomrule
\end{tabular}
\end{table}

Several key observations emerge from the QM7 results. First, unlike the TiO$_2$
experiments, the BNN models (LRT and RAD) achieve lower energy RMSE, even though very similar, than DE on QM7 data
($13.3$ and $13.0$~meV/atom vs.\ $15.2$~meV/atom). This reversal suggests that
the relative performance of Bayesian and ensemble methods is dataset-dependent, and
that the regularization induced by the variational prior may be particularly
beneficial for the more diverse chemical space of QM7.

Second, RAD exhibits again a positive NLL ($1.5$) consistent with its behaviour on TiO$_2$, related to its tendency to predict small uncertainties. LRT
achieves reasonable accuracy but shows high NLL variance ($\pm 2.3$), reflecting
instability in uncertainty calibration across runs.

\begin{figure}[h]
\centering
\includegraphics[width=0.8\textwidth]{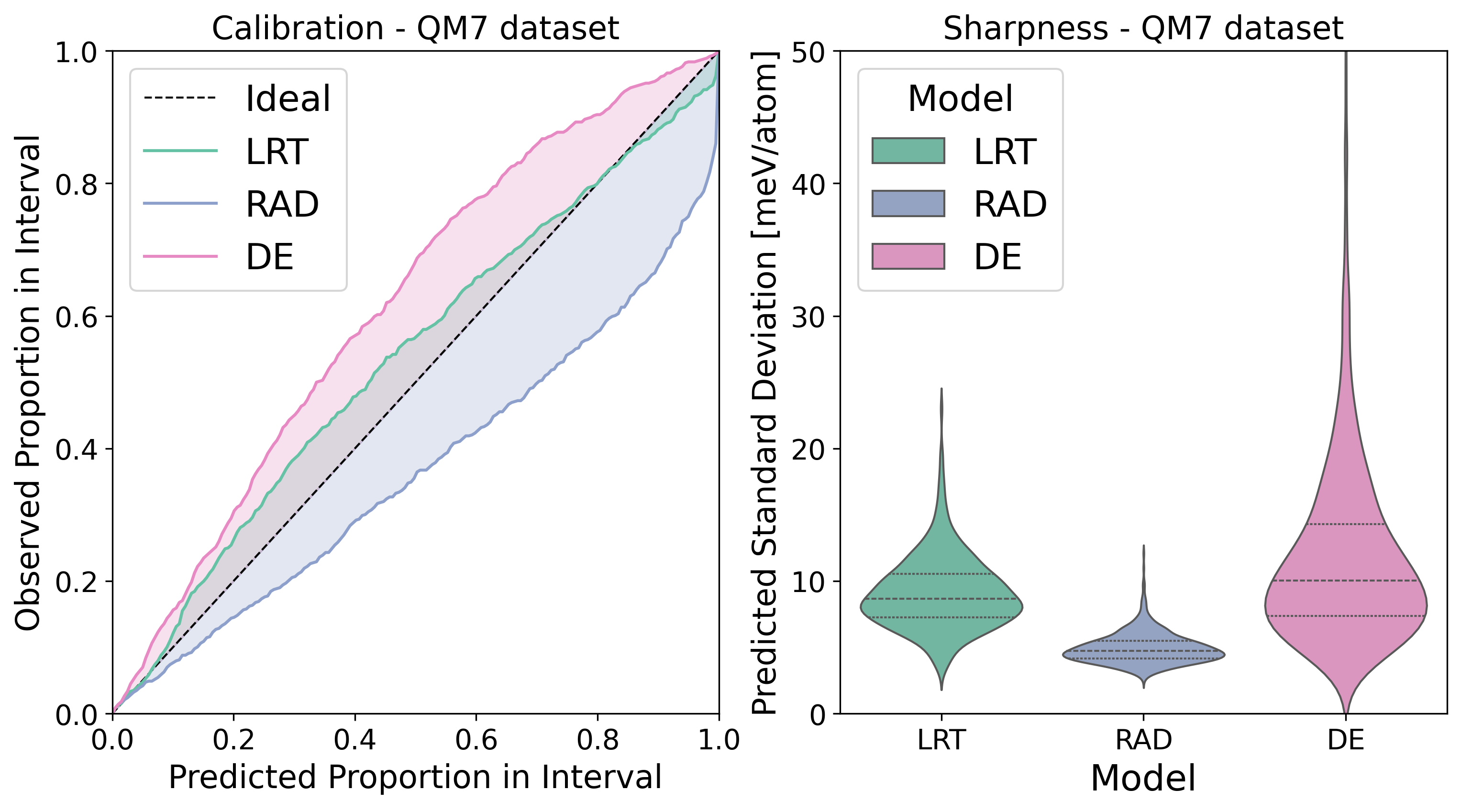}
\caption[QM7 calibration curves]{Calibration curves (left panel) and predicted standard deviations distributions (right panel) for the training with better calibration (lowest RMSCE)  for all the three different models. }
\label{fig:qm7_calibration}
\end{figure}

Figure~\ref{fig:qm7_calibration} shows the calibration curves for all models.
LRT and DE achieve the best calibration (lowest RMSCE), while RAD shows poorer
calibration. These results demonstrate that the conclusions drawn from
TiO$_2$ broadly hold on QM7: LRT provides the most reliable BNN variant across
both predictive accuracy and uncertainty quality.

Our QM7 results thus suggest that chemical diversity primarily affects predictive accuracy, while the relative ordering of uncertainty quantification methods is preserved across datasets.

\subsection{Training and Convergence Time}
\label{sec:results_training_time}

Table~\ref{tab:training_times_stats} presents a comparative overview of training
and inference efficiency across the uncertainty-aware models under both low- and
high-data regimes, benchmarked on a single NVIDIA RTX 5090 GPU using each model's
optimized batch size and number of Monte Carlo samples. At the level of a single
gradient step, DE retains the lowest per-epoch cost in both regimes ($0.066$ and
$0.224$~s in the low- and high-data settings, respectively), thanks to the simple
deterministic forward pass of each member network, while the Bayesian models incur
a higher per-epoch cost driven by the ELBO computation and variational sampling.
However, because a deep ensemble trains $M = 5$ independent networks, its
total training cost is correspondingly higher: in the high-data regime it
reaches $1.3$~h for the full ensemble, comparable to LRT ($1.2$~h) and above RAD
($0.9$~h), while in the low-data regime ($0.2$~h) it lies between RAD ($0.1$~h) and
LRT ($0.4$~h). The
number of epochs to convergence varies strongly with the data regime and the
stochasticity of the Bayesian optimization, with the BNNs requiring more epochs
under data scarcity. Per-epoch times are reported separately for each regime, since
the number of gradient steps per epoch depends on both the dataset size and the
batch size selected by hyperparameter optimization
(Tables~\ref{tab:optuna_hyperparams_20} and~\ref{tab:optuna_hyperparams_100}).

The clearest computational advantage of DE emerges at inference time. Predicting energies and forces on the test set takes only $0.03$~s for the full ensemble, compared with $0.64$~s (LRT) and $0.71$~s (RAD) for the Bayesian models because BNN prediction requires $N_s = 20$ Monte Carlo weight samples, each entailing an additional force evaluation through automatic differentiation, whereas DE needs a single deterministic forward and backward pass per member. Inference cost is regime-independent, as the test set, network architecture, and number of samples are identical across regimes.

\begin{table}[h]
\centering
\caption[Training and inference timing]{Epochs to convergence (mean $\pm$ std over 5 runs), per-epoch wall-clock time, total training time, and inference time for each model in the low- and high-data regimes, measured on a single NVIDIA RTX 5090 using each model's optimized batch size and number of Monte Carlo samples (Tables~\ref{tab:optuna_hyperparams_20} and~\ref{tab:optuna_hyperparams_100}). Total time is (epochs to converge) $\times$ (time per epoch), $\times 5$ for DE (per-epoch time for DE is per ensemble member; the total corresponds to the full 5-member ensemble). Inference is the wall-clock to predict energies and forces on the test set ($N_s = 20$ samples for the BNNs; full ensemble for DE; batch size 128) and is regime-independent.}
\label{tab:training_times_stats}
\begin{tabular}{llcccc}
\toprule
\textbf{Regime} & \textbf{Model} & \textbf{Epochs to Converge}
& \textbf{Time/Epoch (s)} & \textbf{Total Time (h)} & \textbf{Inference (s)} \\
\midrule
{\textbf{Low}}
& LRT & 6,994 $\pm$ 1,128 & 0.182 & 0.35 $\pm$ 0.06 & 0.64 \\
& RAD & 2,533 $\pm$ 385   & 0.143 & 0.10 $\pm$ 0.02 & 0.71 \\
& DE  & 1,777 $\pm$ 237   & 0.066 & 0.16 $\pm$ 0.02 & 0.03 \\
\midrule
{\textbf{High}}
& LRT & 2,641 $\pm$ 156   & 1.584 & 1.16 $\pm$ 0.07 & 0.64 \\
& RAD & 4,254 $\pm$ 263   & 0.767 & 0.91 $\pm$ 0.06 & 0.71 \\
& DE  & 4,185 $\pm$ 149   & 0.224 & 1.30 $\pm$ 0.05 & 0.03 \\
\bottomrule
\end{tabular}
\end{table}

\section{Conclusions}

In this work, we conducted a systematic comparison of different techniques for uncertainty estimation for neural networks in the context of machine learning interatomic potentials. We tested Deep Ensembles (DE) and two variational Bayesian neural network (VBNN) approaches: Local Reparametrization Trick (LRT) and Radial guide (RAD). The evaluation was carried out on the TiO$_2$ dataset with various crystalline phases of the oxide~\citep{Artrith2016} and, to assess generalizability, on the QM7 benchmark of small organic molecules~\citep{blum,rupp2012fast}. We assessed model performance across both high-data and low-data regimes using a comprehensive suite of metrics designed to quantify both predictive accuracy and the quality of uncertainty quantification.

Our findings showed that, regardless of the more empirical and less-theory-grounded approach, DE consistently demonstrated superior performance across almost all metrics and the two data regimes. DE achieved the highest predictive accuracy (lowest MAE and RMSE), the most calibrated uncertainty estimates (lowest RMSCE and NLL), and the most informative uncertainty quantification, as reflected by the highest $r_s$ and overlap scores under data scarcity, though LRT achieved a comparable or slightly higher $r_s$ in the high-data regime. Notably, in the low-data regime DE retained sharp (i.e.\ narrow) uncertainty intervals comparable to RAD while still preserving the strongest correlation between predicted errors and predicted uncertainties, in contrast to LRT, whose sharpness value grew substantially under data scarcity.
Despite its non-Bayesian framework, it is not new that DE empirically outperforms BNNs. Recent work has shown that DE can be interpreted as performing empirical Bayes inference wherein the prior is learned from the data~\citep{Loaiza-Ganem2025} and as a multi-modal sampling of the posterior distribution~\citep{Jospin2020}.

Among the VBNN approaches, LRT offered the best trade-off between accuracy and UQ, showing competitive performance with DE. Both BNN variants (LRT and RAD) achieved notably lower force RMSE than DE under data scarcity, suggesting enhanced robustness to outliers conferred by the variational prior. RAD, while performing comparably to LRT in the high-data regime, showed instability in the low-data setting with elevated energy errors and poorly calibrated uncertainties. Both LRT and RAD showed more pronounced degradation in calibration and uncertainty correlation than DE under low-data conditions.

These findings suggest that ensemble-based methods, such as DE, offer major advantages for uncertainty-aware modelling, especially due to their lower implementation complexity and their low per-step and inference cost. We note, however, that the total training cost of a deep ensemble scales with the number of members and can be comparable to or exceed that of a single variational BNN in the high-data regime, so its computational advantage is most pronounced at inference time rather than during training. VBNNs, on the other hand, offered overall good and comparable predictive performance from a single model, at the price of a substantially higher inference cost. Moreover, the complexity of the VBNN training task makes them more susceptible to random initialization, suggesting that more than one training should be performed to achieve ideal performance. Nonetheless, such higher complexity, related to their stronger theoretical background, still makes them an interesting tool to quantify uncertainty, allowing for a more informed and tunable modelling over the specific dataset.

Evaluation on the chemically more diverse QM7 dataset reveals that these conclusions are not fully universal: while DE retains superior uncertainty quantification across all metrics, the BNN models (LRT and RAD) achieve lower energy RMSE than DE (13.3 and 13.0 versus 15.2~meV/atom), suggesting that the regularisation induced by the variational prior can be particularly beneficial for datasets spanning a broader chemical space. This dataset-dependence of predictive accuracy underscores the importance of validating uncertainty quantification methods across chemically distinct systems before drawing general conclusions.

Taken together, our results suggest a practical recipe: DE remains the preferred choice when reliable uncertainty quantification is the priority and the chemical space is relatively narrow, while VBNN approaches, particularly LRT, become more attractive in data-scarce or chemically diverse regimes where the variational prior provides useful regularisation.

\section*{Acknowledgements}
This work was supported by the Spanish/FEDER
Ministerio de Ciencia, Innovacion y Universidades
[Grant Nos. PID2021-128217NB-I00, MDM-2017-0767, CEX2021-001202-M, PID2022-140120OA-I00, and RYC2021-032281-I (for A.B.)] as well as by the Generalitat de Catalunya [Grant No. 2021SGR00286]. R. F. thanks the Spanish MICIUN for an FPI PhD grant (MDM-2017-0767-20-2). R. F. further acknowledges his current research group for supporting the continuation and development of this work. Computer resources have been partly provided by the Red
Espa$\mathrm{\tilde{n}}$ola de Supercomputacion. This study was also supported by the European COST Actions CA18234 and CA21101.

\section*{Data Availability}

The implemented library (along with documentation) are available in the GitHub repository bayesaenet (https://github.com/farrisric/bayesaenet).

\bibliographystyle{unsrtnat}
\bibliography{references} 

\end{document}


\maketitle

\section*{Code availability}
The code and scripts used to perform the training can be found at~\url{https://github.com/farrisric/bayesaenet}.

\section{Energy-only TiO$_2$ training}
\label{si:sec:energy_only}
The training on the TiO$_2$ dataset was also performed without including the forces, for all the models described in the main text, with the addition of a fourth Bayesian technique, Flipout (FO), which is not compatible with force training (see main text, Section~2.2). The trainings were performed for both low-data and high-data regimes, as described in the main text. For direct comparison with the force-trained results, see Tables~4 and~5 of the main text.

\subsection{Hyperparameters}

Tables~\ref{tabsi:optuna_hyperparams_20} and~\ref{tabsi:optuna_hyperparams_100} report the hyperparameters used for the low-data and high-data regimes, respectively.

\begin{table}[h]
\centering
\caption[Energy-only hyperparameters, low-data regime]{Best hyperparameters selected via Optuna for each model trained on 20\% of the TiO$_2$ dataset (energy-only training).}
\label{tabsi:optuna_hyperparams_20}
\begin{tabular}{lccc}
\toprule
\textbf{Hyperparameter} & \textbf{FO} & \textbf{LRT} & \textbf{RAD} \\
\midrule
Learning Rate      & $3.25\times10^{-4}$ & $5.40\times10^{-4}$ & $1.37\times10^{-4}$ \\
Batch Size         & 64 & 64 & 32 \\
MC Samples         & 2 & 2 & 2 \\
Prior Scale        & 0.175 & 0.358 & 0.115 \\
$q_\theta$ Scale   & $1.83\times10^{-3}$ & $1.25\times10^{-3}$ & $1.72\times10^{-4}$ \\
$\sigma_E$         & 0.260 & 0.282 & 0.793 \\
\bottomrule
\end{tabular}
\end{table}

\begin{table}[h]
\centering
\caption[Energy-only hyperparameters, high-data regime]{Best hyperparameters selected via Optuna for each model trained on 100\% of the TiO$_2$ dataset (energy-only training).}
\label{tabsi:optuna_hyperparams_100}
\begin{tabular}{lccc}
\toprule
\textbf{Hyperparameter} & \textbf{FO} & \textbf{LRT} & \textbf{RAD} \\
\midrule
Learning Rate      & $1.24\times10^{-4}$ & $4.90\times10^{-5}$ & $5.36\times10^{-4}$ \\
Batch Size         & 256 & 64 & 64 \\
MC Samples         & 2 & 2 & 2 \\
Prior Scale        & 0.206 & 0.209 & 0.108 \\
$q_\theta$ Scale   & $6.05\times10^{-4}$ & $2.27\times10^{-4}$ & $8.00\times10^{-4}$ \\
$\sigma_E$         & 0.893 & 0.132 & 0.294 \\
\bottomrule
\end{tabular}
\end{table}

\subsection{Accuracy and uncertainty quantification}

\begin{figure}[h]
\centering
\includegraphics[width=\textwidth]{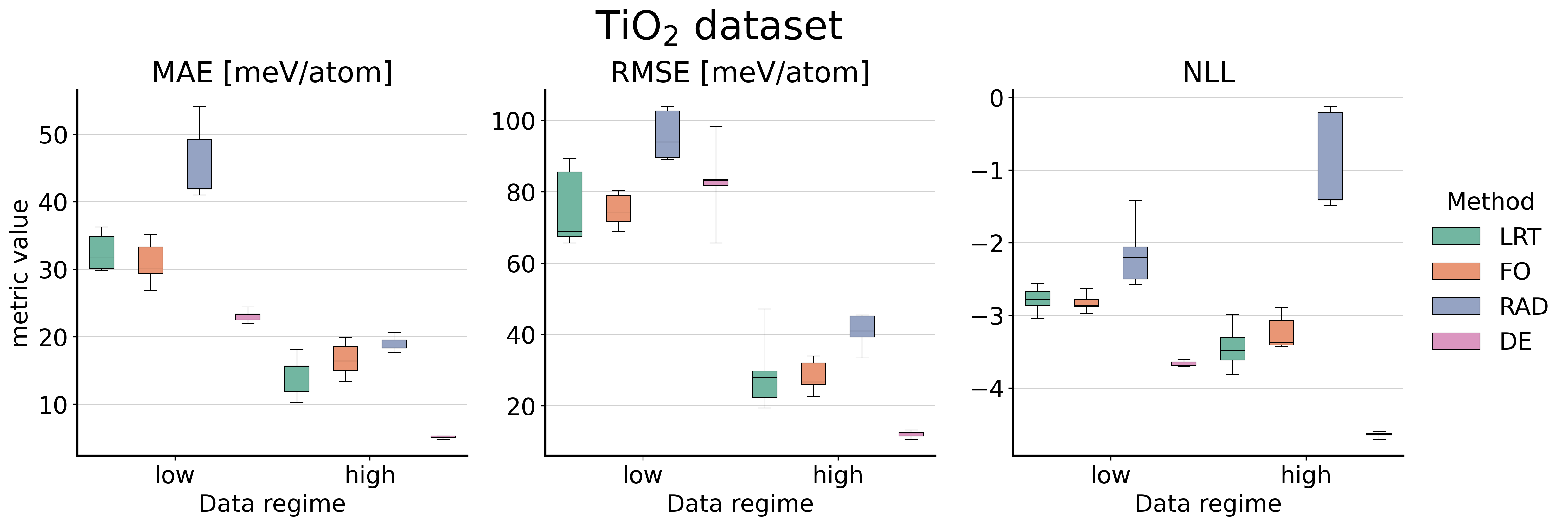}
\caption[Energy-only accuracy metrics, TiO$_2$]{Predictive performance metrics (MAE, RMSE, NLL) for all models trained on TiO$_2$ in the energy-only setting, across high-data (100\% of the dataset) and low-data (20\% of the dataset) regimes. Boxplots summarize the distribution of scores over five independent training runs: the boxes show the interquartile range (25th to 75th percentile), the horizontal line within each box marks the median, and the whiskers extend to the whole range of the scores. Colour coding: LRT (green), FO (orange), RAD (blue) and DE (pink). The associated values are reported in Table~\ref{tabsi:avg_metrics}.}
\label{figsi:performance_de_bnn}
\end{figure}

\begin{figure}[h]
\centering
\includegraphics[width=\textwidth]{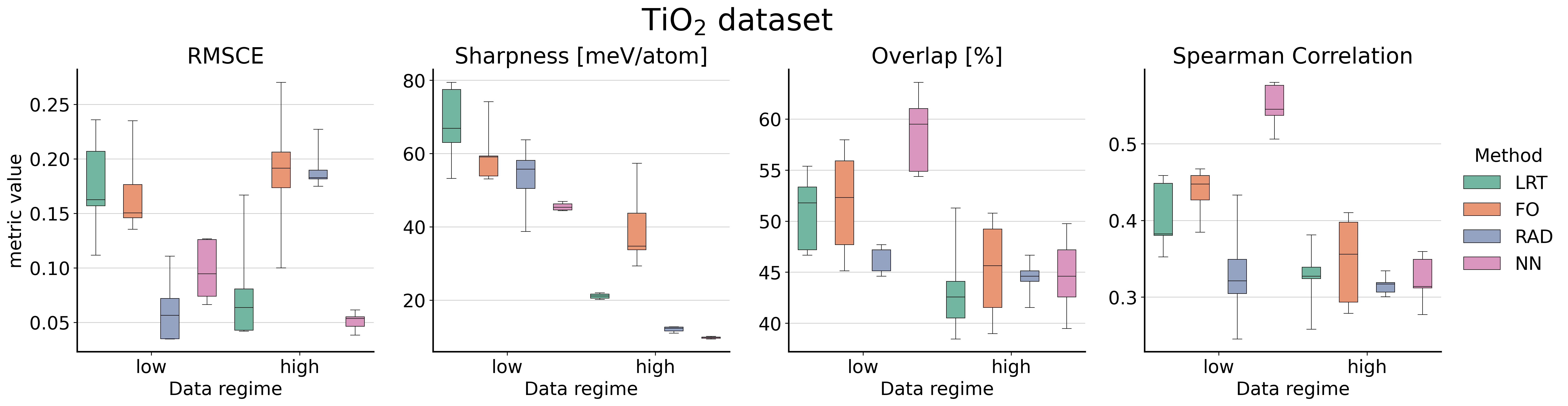}
\caption[Energy-only UQ metrics, TiO$_2$]{Root mean squared calibration error (RMSCE), sharpness, overlap and Spearman correlation coefficient ($r_s$) of the predicted uncertainty distributions across methods and data regimes for the energy-only TiO$_2$ training. Boxplots summarize the distribution of scores over five independent training runs: the boxes show the interquartile range (25th to 75th percentile), the horizontal line within each box marks the median, and the whiskers extend to the whole range of the scores. Colour coding: LRT (green), FO (orange), RAD (blue) and DE (pink). The associated values are reported in Table~\ref{tabsi:uq_metrics}.}
\label{figsi:uq_quality}
\end{figure}

\begin{figure}[h]
    \centering
    \includegraphics[width=0.8\linewidth]{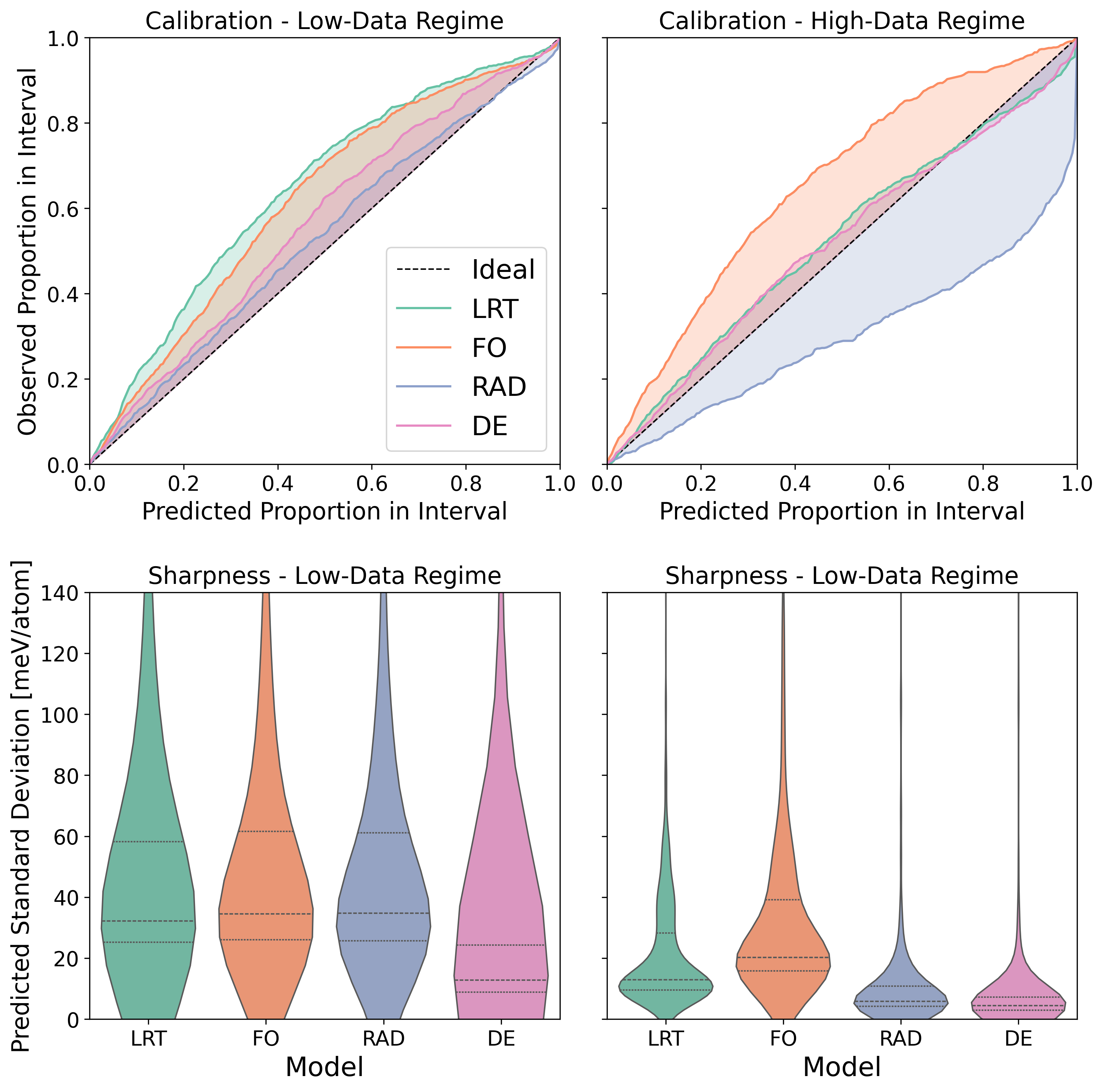}
    \caption[Energy-only calibration and sharpness, TiO$_2$]{Calibration curves (top) and sharpness distributions (bottom) for the best-performing models in each data regime in the energy-only TiO$_2$ training, selected based on the lowest RMSCE value.}
    \label{figsi:cal_sharp_best}
\end{figure}

Figures~\ref{figsi:performance_de_bnn},~\ref{figsi:uq_quality} and~\ref{figsi:cal_sharp_best} show the metrics used to quantify the quality of the predictions and uncertainties obtained, following the same conventions as in the main text. The numeric values are reported in Tables~\ref{tabsi:avg_metrics} and~\ref{tabsi:uq_metrics}. Overall, FO is competitive with LRT in both energy accuracy and calibration, and without forces all models reach lower energy MAE than in the joint-likelihood training reported in the main text, while the relative ordering across methods is preserved.

\begin{table}[h]
\centering
\caption[Energy-only TiO$_2$ accuracy metrics]{Average MAE, RMSE, and NLL for each model trained on the TiO$_2$ dataset without forces, under both high-data and low-data regimes. Energy metrics are reported in meV/atom; uncertainty values indicate the standard deviation over five independent runs.}
\label{tabsi:avg_metrics}
\begin{tabular}{llccc}
\toprule
\textbf{Regime} & \textbf{Method} 
& \textbf{MAE [meV/atom]} 
& \textbf{RMSE [meV/atom]} 
& \textbf{NLL} \\
\midrule

\textbf{High-data}
& LRT & $14 \pm 3$  & $29 \pm 11$ & $-3.44 \pm 0.31$ \\
& FO  & $17 \pm 3$  & $28 \pm 5$  & $-3.24 \pm 0.24$ \\
& RAD & $19 \pm 1$  & $41 \pm 5$  & $-0.93 \pm 0.69$ \\
& DE  & $5 \pm 1$   & $12 \pm 1$  & $-4.65 \pm 0.04$ \\
\midrule

\textbf{Low-data}
& LRT & $33 \pm 3$  & $75 \pm 11$ & $-2.79 \pm 0.18$ \\
& FO  & $31 \pm 3$  & $75 \pm 5$  & $-2.83 \pm 0.13$ \\
& RAD & $46 \pm 6$  & $96 \pm 7$  & $-2.15 \pm 0.46$ \\
& DE  & $23 \pm 1$  & $82 \pm 12$ & $-3.67 \pm 0.04$ \\
\bottomrule
\end{tabular}
\end{table}

\begin{table}[h]
\centering
\caption[Energy-only TiO$_2$ UQ metrics]{Average RMSCE, Sharpness, Overlap, and Spearman correlation (r$_s$) for each model trained on the TiO$_2$ dataset without forces, under both high-data and low-data regimes. Uncertainty values indicate the standard deviation over five independent runs.}
\label{tabsi:uq_metrics}
\begin{tabular}{llcccc}
\toprule
\textbf{Regime} & \textbf{Method} 
& \textbf{RMSCE} 
& \textbf{Sharpness [meV/atom]} 
& \textbf{Overlap [\%]} 
& \textbf{r$_s$} \\
\midrule

\textbf{High-data}
& LRT & $0.08 \pm 0.05$ & $21 \pm 1$  & $43 \pm 5$ & $0.33 \pm 0.04$ \\
& FO  & $0.19 \pm 0.06$ & $40 \pm 11$  & $45 \pm 5$ & $0.35 \pm 0.06$ \\
& RAD & $0.19 \pm 0.02$ & $12 \pm 1$  & $44 \pm 2$ & $0.32 \pm 0.01$ \\
& DE  & $0.05 \pm 0.01$ & $10 \pm 1$  & $45 \pm 4$ & $0.32 \pm 0.03$ \\
\midrule

\textbf{Low-data}
& LRT & $0.17 \pm 0.05$ & $68 \pm 11$ & $51 \pm 4$ & $0.40 \pm 0.05$ \\
& FO  & $0.17 \pm 0.04$ & $60 \pm 8$ & $52 \pm 5$ & $0.44 \pm 0.03$ \\
& RAD & $0.06 \pm 0.03$ & $53 \pm 9$  & $46 \pm 1$ & $0.33 \pm 0.07$ \\
& DE  & $0.10 \pm 0.03$ & $45 \pm 1$ & $59 \pm 4$ & $0.55 \pm 0.03$ \\
\bottomrule
\end{tabular}
\end{table}
